\documentclass[fleqn,10pt]{wlscirep}
\usepackage[utf8]{inputenc}
\usepackage[T1]{fontenc}
\usepackage{amsmath,amsfonts}
\usepackage{algorithmic}
\usepackage{array}
\usepackage[caption=false,font=normalsize,labelfont=sf,textfont=sf]{subfig}
\usepackage{textcomp}
\usepackage{stfloats}
\usepackage{url}
\usepackage{verbatim}
\usepackage{graphicx}
\usepackage{cite}
\usepackage{mathtools}
\usepackage{tabularx}
\usepackage{multirow}
\usepackage{booktabs}
\usepackage[version=4]{mhchem} 
\usepackage{siunitx}
\usepackage[ruled, lined, linesnumbered, commentsnumbered, longend]{algorithm2e}
\usepackage{setspace}
\usepackage{xcolor}
\usepackage{subfig}

\SetCommentSty{mycommfont}

\SetKwInOut{KwIn}{Input}
\SetKwInOut{KwOut}{Output}
\SetKwFor{While}{while}{}{end while}%
\SetKwRepeat{Do}{do}{while}

\let\oldnl\nl% Store \nl in \oldnl
\newcommand{\nonl}{\renewcommand{\nl}{\let\nl\oldnl}}% Remove line number for one line

\newlength\mylen
\newcommand\myinput[1]{%
  \nonl
  \settowidth\mylen{\KwIn{}}%
  \setlength\hangindent{\mylen}%
  \hspace*{\mylen}#1
  \\ }
  
\newcommand{\doublehat}[1]{% 
\begingroup%
  \let\macc@kerna\z@%
  \let\macc@kernb\z@%
  \let\macc@nucleus\@empty%
  \hat{\raisebox{.2ex}{\vphantom{\ensuremath{#1}}}\smash{\hat{#1}}}%
\endgroup%
}
\makeatother

\newcommand\sza{0.208}
\newcommand\szb{0.08em}

\DeclareMathOperator*{\argmax}{arg\,max}
\DeclareMathOperator*{\argmin}{arg\,min}

\newcolumntype{x}[1]{>{\centering\arraybackslash\hspace{0pt}}p{#1}}

\title{A Deep Generative Approach to Oversampling in Ptychography}

\author[1,*]{Semih Barutcu}
\author[1]{Aggelos K. Katsaggelos}
\author[1,2]{Do\u ga G\" ursoy}
\affil[1]{Northwestern University, 2145 Sheridan Road, Evanston, IL, 60208, USA}
\affil[2]{Argonne National Laboratory, 9700 South Cass Avenue, Lemont, IL, 60439, USA}
\affil[*]{semihbarutcu@u.northwestern.edu}

\begin{abstract}

Ptychography is a well-studied phase imaging method that makes non-invasive imaging possible at a nanometer scale. It has developed into a mainstream technique with various applications across a range of areas such as material science or the defense industry. One major drawback of ptychography is the long data acquisition time due to the high overlap requirement between adjacent illumination areas to achieve a reasonable reconstruction. Traditional approaches with reduced overlap between scanning areas result in reconstructions with artifacts. In this paper, we propose complementing sparsely acquired or undersampled data with data sampled from a deep generative network to satisfy the oversampling requirement in ptychography. Because the deep generative network is pre-trained and its output can be computed as we collect data, the experimental data and the time to acquire the data can be reduced. We validate the method by presenting the reconstruction quality compared to the previously proposed and traditional approaches and comment on the strengths and drawbacks of the proposed approach.

\end{abstract}

\begin{document}

\flushbottom
\maketitle
\thispagestyle{empty}

\section{Introduction}

Coherent Diffraction Imaging (CDI) is a collection of lens-free techniques to achieve nanometer-scale reconstructions by avoiding lens-based limitations of microscopy \cite{Abbey2008, Dierolf2008, Chapman2010, Miao2012, Vine2012, Miao2015}. Ptychography, a scanning CDI technique, is a method to achieve non-invasive imaging of large field-of-views with sub-\SI{10}{\micro\meter} spatial resolution \cite{gursoy2021lensless}. In ptychography, a focused, coherent beam of light called probe scans the object in an overlapping trajectory, and the intensities of the resulting diffraction patterns are recorded at the far-field detector \cite{Fienup:78}. The scanning procedure enables imaging larger field of views than the focused beam size. Similarly, without a lens, we avoid lens-induced distortions and artifacts caused by low photon efficiency \cite{Adams2015}. However, the mechanical scanning procedure also results in increased time for obtaining the images compared to conventional x-ray microscopy imaging methods. Although the scanning process can be optimized, high-resolution reconstructions or more complex models require lengthened data acquisition process. For instance, scanning time increases significantly in 3D ptychography, where the object is scanned from different view-angles for future tomographic reconstructions \cite{PhysRevA.99.023801, Barutcu2020, du2021upscaling}.

High data acquisition time in ptychography leads to a trade-off between data acquisition methods. The trade-off occurs between high photon count and high overlap percentage for a given acquisition time. To keep the acquisition time low, the dwelling time for each scan position can be lowered, leading to a low photon count. This increases the noise in the measured intensities \cite{Savikhin2019}. On the contrary, the overlap percentage can be reduced by decreasing the number of scan points while increasing the exposure time per scan point to have few but more accurate images. In many cases, avoiding the random noise is challenging, leading to the increased importance of quality reconstruction with a low amount of overlapping data\cite{osti_1599580_ptychnet}. Apart from the restricted data acquisition time, increased reconstruction quality with a low overlap percentage can be crucial for critical material science studies and for imaging larger volumes \cite{Shamshad_Ahmed_2018}.

Traditional methods solve ptychographic phase retrieval problems with different approaches. Wigner deconvolution method converts the phase retrieval process to a blind deconvolution problem \cite{Wigner1992}. Difference Map (DM) converts the inverse problem into a constrained optimization problem and solves it iteratively \cite{Thibault_Dierolf_Menzel_Bunk_David_Pfeiffer_2008a, thibault2009probe, du2020three}. Similar to the DM, the extended Ptychographical Iterative Engine (ePIE) technique recovers the phase by updating the complex object's magnitude at each iteration\cite{thibault2009probe}. These methods and their improved variations generate high-resolution reconstructions with a low amount of artifacts when there is a high overlap percentage between the scanning points, the noise level is low, and the scan process details are known with high precision \cite{bouman}. However, the data acquisition time required for such experiments is high and traditional methods fail to recover the lost information when there is low overlap between adjacent scans\cite{thibault2009probe}.

To lower the required data acquisition time, one possible option is replacing stop-and-shoot type data acquisition with constant-velocity scanning \cite{Pelz:14, Deng:15a, Clark:14}. However, constant-velocity scanning is challenging due to the stability and control of the stages during mechanical scanning. There have been developments focusing on solving the compressed sensing problem where there is insufficient redundancy in the acquired data to recover the lost phase information \cite{Candes:06, Donoho:06}. For instance, removing some of the random pixels of the detector to lower the required data\cite{Stevens2018_subsamp_ptycho}, Gabor's decomposition to solve the inverse problem \cite{daSilva2015_decomp}, and applying lifting techniques and iterative reweighing \cite{Moravec2007_pr_theo1, Ohlsson2012_pr_theo2, Newton2012_pr_theo3} are possible applications of compressed sensing to ptychography. However, while these methods provide some insights on possible applications, they provide marginal reductions in the data acquisition time. Removal of the overlap requirement altogether, which has the highest impact on the time required for the data acquisition process, or reconstructions under significantly noisy conditions, which can be a situation where the mechanical scanning is done too fast, remain largely unsolved. 

Deep learning brought a new point of view to various problems in computational imaging literature. Phase retrieval problems are also reconsidered by utilizing different deep learning methodologies either as stand-alone solutions or as combinations with well-established techniques \cite{Sinha2017, Cherukara2018, Xue2019, zhou2020diffraction, Barbas2, Barbas3, wohl}. The application of deep generative priors (DGPs) is shown in supervised learning approaches for ptychography \cite{Nguyen2018-cGAN, osti_1599580_ptychnet}. Generative priors are also shown to effectively reduce artifacts caused by imperfect experimental setups \cite{condmat6040036} and lowered overlap percentage in ptychography scans \cite{Aslan_Liu_Nikitin_Bicer_Leyffer_Gursoy_2021, Pan_Zhan_Dai_Lin_Loy_Luo_2021}, which shows the promise for reducing overlap constraint. However, the possibility of achieving high-quality reconstructions is hindered by the requirement for an extensive training dataset for supervised learning. On the other hand, Deep Image Priors (DIPs) method \cite{Ulyanov_Vedaldi_Lempitsky_2020}, which is based on self-optimization rather than training, is shown to be effective for compressed sensing and phase retrieval problems \cite{Shamshad_Ahmed_2018, VanRullen_Reddy_2019a, Shamshad_Ahmed_2021, Barbas1}, and it has been applied to x-ray ptychography with successful results \cite{Yang2020}. However, both DGPs and DIPs are not successful enough to remove the overlapping requirement for ptychography. DGPs require the generator network to be trained such that the solution is required to lie in the representation span of the network. Finding the perfect dataset in nano-scale imaging is challenging, and even with the optimal dataset, the reconstructions are sub-optimal. Similarly, a good reconstruction using optimization with DIPs depends heavily on the initialization of the network weights, otherwise, optimization time increases significantly, and the convergence is challenging. 

In \cite{Barutcu22}, we previously proposed a method to combine DGPs and DIPs in a single framework. In this method, generative networks are pre-trained with a large dataset that closely represents the target reconstruction. The network weights are initialized for the DIPs framework using the network and the DGPs that are learned. The input latent vector is modified for the specific solution, and then the network weights are optimized progressively to extend the reconstruction outside of the representation capability of the learned DGPs. The method also adds external information and the discriminator network information as priors to constrain the solution. These are shown to be total variation prior promoting piece-wise smoothness and discriminator prior punishing the deviation from the training dataset. The network is shown to be superior to traditional methods for reconstructions with low or no overlap simulation data. It is also shown that the noise is handled significantly better than the previous approaches. However, when the missing data in the compressed sensing problem is not extreme, fine details cannot be recovered fully due to the high non-convexity of the inverse problem \cite{Barutcu22}. This situation is also visible when there is enough overlap for traditional methods to have successful reconstructions. Although both approaches produce good results, traditional approaches may recover fine details better \cite{He2011}.

In this paper, we propose an algorithm that combines the traditional methods and the algorithm proposed in our previous work \cite{Barutcu22}. In the proposed approach, we suggest starting with the previously proposed method to get an initial reconstruction for low or no overlap data. The reconstruction is not expected to be perfect, but it will be able to reconstruct the object where the traditional methods fail. Then, using the reconstruction, we apply the forward ptychography model with a much higher overlap percentage, which will result in significantly more data points, including the artificial diffraction data. Then, we suggest applying the ePIE algorithm using the acquired simulated diffraction data, but we replace the data with the original low overlap data for the corresponding scan positions where they exist. Successful reconstructions can be achieved using traditional approaches such as ePIE using this combination at this step. Furthermore, the application of the last step can be repeated multiple times to increase the fidelity to the data.

We validate our method with numerical tests simulating the compressed sensing x-ray ptychography experiment and compare the reconstructions with ePIE and the state-of-the-art method suggested in \cite{Barutcu22}. Moreover, we suggest that the proposed method can also be effective in noisy data by recovering some structural details lacking in similar methods.  Then, we show that the reconstruction quality is further improved by repeating the last step of the algorithm on the same data. Finally, we argue that either the reconstruction quality is increased or artificial artifacts are removed, which are brought by DGPs, resulting in quality reconstructions with no hard overlap constraints in x-ray ptychography.

\section{Method}

In this section, we provide background information on the ptychography phase-retrieval problem, present an approach to improve a state-of-the-art method in the literature, and describe the algorithm that we propose.

\subsection{The Forward Model}
\begin{figure}[t]
    \centering
    \hspace{40pt}
    \includegraphics[width=0.5\columnwidth]{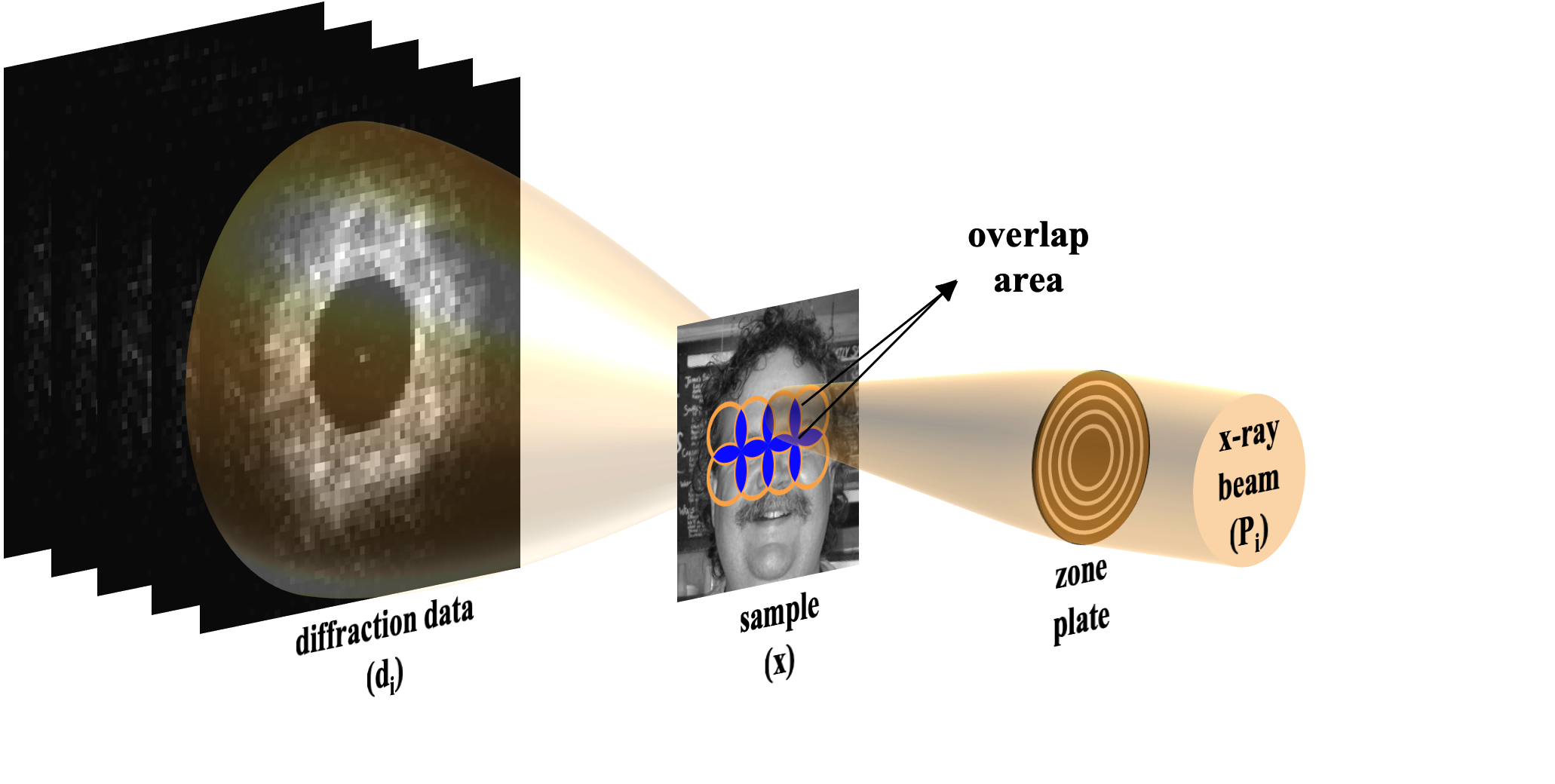}
    \vspace{-5pt}
    \caption{The model for the x-ray ptychography data acquisition setup. A coherent beam probe is scanned across an object where adjacent scan positions (shown in orange) have overlapping regions between them (shown in blue). The intensities of the resulting diffraction patterns are recorded at the detector for each scan position $i$.}
    \label{fig:ptycho}
\end{figure}
In ptychography, an object is scanned with a focused, coherent x-ray beam in an overlapping trajectory between the consecutive scanning positions, and the intensity values are recorded at the pixelated photon-counting detector. The process described can be modeled as in Fig.~\ref{fig:ptycho}. 

We can describe the ptychography acquisition process and the interactions between materials mathematically. The focused x-ray beam (probe) can be described with a 2D complex illumination function at the wave-front denoted by $P \in \mathbb{C}^{S\times S}$. The focused probe interacts with the complex 2D object $x \in \mathbb{C}^{T\times T}$ at position $i$. The size of the probe is considerably smaller than the size of the complex object ($S\leq T$), and the position $i$ denotes a part of the object that is of the same size as the probe. The resulting wave function $\psi \in \mathbb{C}^{S\times S}$ can be shown as:
\begin{equation}\label{Eq:forward1}
     \psi_i = P_i \odot x
\end{equation} 
where $\odot$ denotes element-wise multiplication. The resulting wave ($\psi_i$) goes through far-field Fraunhofer diffraction, and it can be represented by the Fourier transform operation. Then the intensity of the transformed wave can be recorded at the detector, where the phase information is lost. This process can be shown as
\begin{equation}
\label{Eq:forward2}
     d_i = |F \psi_i|^2 + \eta_i
\end{equation} 
where $d_i \in \mathbb{R}^{S\times S}$ is the intensity of the transform detected, $\eta_i$ is the noise added at scan position $i$, $F \in \mathbb{C}^{S\times S}$ represents the Fourier Transform, and $|.|^2$ is the intensity operator.

The complex diffraction signal $F \psi_i$ in Eq.~\ref{Eq:forward2} can be expressed as $A(k)exp(j\Phi(k))$. The detector can only record the intensity values $A(k)$ leading to a loss of information in the phase $\Phi(k)$. This creates an ill-posedness in the inverse problem to recover the phase information. Algorithms suggested for the recovery of the phase information are called phase retrieval algorithms which try recovery by iterative approaches. In many situations, there are some limitations to the complex object to be recovered in the suggested algorithms, such as the object being finite in extent or objects to consist phase information only \cite{gerchberg1972practical, fienup1978reconstruction, fienup1982phase, fienup1987reconstruction}. To avoid these drawbacks, the ptychographic method is developed. To recover the phase in ptychography, the redundant information in the overlapping regions in the forward model is utilized. 

\subsection{The Inverse Problem}

The inverse problem for ptychography can be reconstructed as a minimization problem where the object is recovered as the output. An L1 minimization can be expressed as:
\begin{equation}
\label{Eq:L1}
      \hat{x} = \argmin_{x} \sum_i \| d_i - |F(P_i \odot x)|^2\|_1.
\end{equation} 
where $\hat{x}$ denotes the reconstructed object. The inverse problem is solved using the redundant information in the overlapping regions via iterative approaches traditionally.

Algorithms can be developed to lower or remove the required overlap percentage to reconstruct the object from a lower amount of data than required. The method to recover a signal $m \in \mathbb{C}^{G}$ from forward model $v = Bm \in \mathbb{C}^{H}$ where $G>>H$ and the forward operator is $E \in \mathbb{C}^{H\times G}$ is called compressed sensing. It is the name for methods that use compression and acquisition simultaneously to lower the acquisition time and increase the efficiency of operations. Ptychography with low among of overlap can be considered a compressed sensing problem. To recover signals from compressed data, prior information is required.

Total Variation (TV) regularization is a method to apply strong prior information. It promotes piece-wise continuity, a common feature of artificial objects commonly used as samples of x-ray ptychography. In addition, TV prior can eliminate the possible effects of noisy data by punishing sparse artifacts in the reconstructed object. Thus, TV can be used as a valuable prior to increase reconstruction quality. For a 2D object $x$, total variation distance can be defined as:
\begin{equation}
\label{Eq:tv_reg_x}
      TV(x) = \| \nabla (x) \|_1
\end{equation} 
where $\nabla (x)$ is the 2D gradient of the object. To increase reconstruction quality, the minimization problem in \ref{Eq:L1} can be modified as follows:
\begin{equation}
\label{Eq:tv_reg_x_added}
      \hat{x} = \argmin_{x} \sum_i  \| d_i - |F(P_i \odot x)|^2\|_1 + \lambda_1 \| \nabla (x) \|_1.
\end{equation}
where the regularization parameter $\lambda_1$ is a hyper-parameter, managing the trade-off between the regularization and the data fidelity terms. 

Although the minimization in Eq.~\ref{Eq:tv_reg_x_added} works well for additive noise, the photon counts at the detector can be best described as a Poisson process. Therefore, if we assume that the measurements are independent of each other, we can express the probability of measuring each data point by the likelihood function to maximize the probability of measuring all data.
\begin{eqnarray}
\label{Eq:d_given_x}
    \hat{x} &=& \argmax_{x} p(d|x) \\
    &=& \argmax_{x} \prod_{i} \frac{e^{|F(P_i \odot x)|^2} |F(P_i \odot x|^{2d_i}}{d_i!}
\end{eqnarray} 
Assuming we gave prior information on the object to be reconstructed, we can maximize the a posteriori probability $p(x|d)$ using Bayes' theorem. If we also select the object in the Gibbs form for the piece-wise continuity $p(x) = e^{\lambda_1 \| \nabla (x) \|_1}$, we can convert the minimization problem to a maximization of a posteriori probability as:
\begin{equation}
\label{Eq:map}
    \hat{x} = \argmin_{x} \sum_{i} (|F(P_i \odot x)|^2 - 2d_i log|F(P_i \odot x)|) + \lambda_1 \| \nabla (x) \|_1
\end{equation}
which is more suited for the Poisson process of x-ray ptychography.

\subsection{PtychoGAN Solution}

To address the compressed sensing problem in ptychography, a novel approach has been proposed in \cite{Barutcu22}. Deep Generative Priors (DGPs), Deep Image Priors (DIPs), TV prior, and discriminative prior are combined to get a successful reconstruction from low or no overlap scan data. 

DGPs are utilized as a good initialization method for the following DIPs method that is proposed. When it is trained with a good training dataset, a generator network of a Generative Adversarial Network (GAN) \cite{goodfellow2014generative} structure learns a mapping from a low dimensional latent vector $z \in \mathbb{R}^{k}$ to a higher dimensional representation space $G(z) \in \mathbb{R}^{TxT}$ where $TxT >> k$. Depending on the training dataset and its spanning capability of the object space, the representation capability of the trained generator changes. In the suggested method, the network is trained with a dataset consisting of images similar to the target reconstruction. Then, the minimization is performed to find the best latent vector $z$ to achieve an initial reconstruction. 

After the initial reconstruction, DIPs are used to get the final reconstruction. Even though deep networks are shown to be effective in various situations, their effectiveness is fundamentally limited by the training data used. To achieve reconstructions outside of the representation capability of the generator networks, the low-dimensional image representation capability of deep neural networks can be leveraged to fit the reconstructions to the measurement data. The proposed architecture in \cite{Barutcu22} uses DGPs to fine-tune the network weights to achieve the reconstructions, using the network structure as a prior to lower the requirement for the overlaps. However, optimizing the weights of a large neural network is not easy  due to the high non-convexity of the minimization. To overcome this challenge, the proposed architecture uses progressive adjustment of the network weights, where the updates of the weights start from the shallowest layers, and other weights are added to the minimization sequentially. This allows the minimization to avoid some possible local minima.

In addition to the DGPs and DIPs, the proposed algorithm uses two external priors. TV prior is used to promote sparsity in the final reconstruction. Also, discriminator prior is created to avoid divergence from the training dataset. While the generator network is trained, it is coupled with a discriminator network which learns to decide if the output of the generator network belongs to the desired object space. Therefore, a prior punishing the divergence from the desired domain while updating weights is used for a better reconstruction. 

If the generator network is denoted as $G_{\theta}(.)$ where $\theta$ shows the weights of the network, the discriminator network is shown by $D(.)$, and $z$ is the input latent vector, the minimization problem proposed in \cite{Barutcu22} can be written as:
\begin{equation}
\label{Eq:converted}
    \hat{z}, \hat{\theta} = \argmin_{z, \theta} \sum_i \| |F(P_i \odot e^{jG_{\theta}(z)}|^2 - 2d_i log|F(P_i \odot e^{jG_{\theta}(z)}|\|_1 + \lambda_1 \| \nabla (G_{\theta}(z)) \|_1 + \lambda_2 log (1 - D((G_{\theta}(z)).
\end{equation}
In Eq.~\ref{Eq:converted}, $\lambda_1$ controls the TV prior, $\lambda_2$ controls the discriminative prior, and the minimization is done in a progressive approach as described. The results are shown to be effective for very low or no overlap data. 

\subsection{ePIE Solution}

One of the most widely used algorithms to solve the ptychography phase retrieval problem is extended Ptychography Iterative Engine (ePIE), and it is shown to be successful provided that there is enough overlap data \cite{Maiden2009}. Although the proposed algorithm provides update functions for both the probe $P$ and the object $x$, we will assume that we have a known probe and describe the update on the object only.

The method is an iterative update approach where the object is initialized as an empty array $x^0$ where the superscript shows the iteration number, and the scanned areas on the objects are updated iteratively until convergence. From the initial guess of the object, an initial guess at the exit wave is formed. This process is repeated at every $k^th$ step with:
\begin{equation}
\label{Eq:epie1}
    \psi_i^k = P_i \odot x^k
\end{equation} 
The Fourier transform of the resulting wave function guess is taken, and the magnitude is replaced by the positive square root of the corresponding diffraction pattern. The updated exit wave is calculated by taking the inverse Fourier transform of the value. If the inverse Fourier transform is denoted by $F^{-1} \in \mathbb{C}^{S\times S}$, and the updated exit-wave functions is denoted by $\psi^{k+1} \in \mathbb{C}^{S\times S}$:
\begin{equation}
\label{Eq:epie2}
    \psi_i^{k+1} = F^{-1} \{\sqrt{d_i} \frac{F\psi_i^k}{|F\psi_i^k|}\}
\end{equation} 
Using the updated exit-wave function, the object can be updated. For the update, the difference between the updated exit-wave function and the previous version is taken, and multiplied with the conjugate of the probe function $P^* \in \mathbb{C}^{S\times S}$, and divided by the maximum value of the square of the magnitude of the probe to get the update on the object. Then this update is added to the previous iteration of the object to find the new object guess
\begin{equation}
\label{Eq:epie3}
    x^{k+1} = x_{k} + \alpha \frac{P^*_i}{|P_i|^2_{max}} (\psi_i^{k+1} - \psi_i^{k})
\end{equation} 
where $\alpha$ is a constant to adjust the step size of the update if needed. This process continues with all diffraction patterns, at which point a single ePIE iteration has been completed. Then these iterations are repeated until convergence for the final reconstruction. 
\subsection{Proposed Solution}

\begin{algorithm} [tb]
\setstretch{1}
\caption{Proposed Solution}
\label{alg:alg1}

\KwIn{$d_m \in \mathbb{R}^{S\times S}$, intensity of diffractions at pos. \textit{i}}
\myinput{$z \in \mathbb{R}^{k}$, a random vector}
\myinput{$x_{prop}^{0}$, empty matrix for reconstruction initialization}
\KwOut{$\hat{x} \in \mathbb{R}^{NxN}$, reconstructed object}

\nonl \tcp{Find an initial reconstruction using PtychoGAN from M diffraction patterns}
\nl \hspace{2pt} $\hat{z}, \hat{\theta} = \argmin_{z, \theta} \sum_m \| |F(P_ \odot e^{jG_{\theta}(z)}|^2 - 2d_{m, low} log|F(P_m \odot e^{jG_{\theta}(z)}|\|_1 + \lambda_1 \| \nabla (G_{\theta}(z)) \|_1 + \lambda_2 log (1 - D((G_{\theta}(z))$
\\ $\hat{x}_{low} = e^{G_{\hat{\theta}}(\hat{z})}$
\nonl \tcp{Apply forward model on $\hat{x}_{low}$ to get simulated data with $N = Y^2 M$ where Y is the oversampling ratio}
\nl $d_{n, high} = |F (P_n \odot \hat{x}_{low})|^2$
\nonl \tcp{Combine real and simulated diffraction data}
\nl $d_{n, mixed} \gets d_{m, low}, d_{n, high}$
\nonl \tcp{Apply ePIE to get final reconstruction}
\nl \While {Not converged}
{
    \nl \For{$n \gets 1$ to $N$}
    {
         ${x_{prop}^{k+1} = x_{prop}^{k} + \alpha \frac{P^*_n}{|P_n|^2_{max}} \{(F^{-1} \{\sqrt{d_{n, mixed}} \frac{F P_n \odot x_{prop}^{k}}{|F P_n \odot x_{prop}^{k}|}\}) - (P_n \odot x_{prop}^{k})\}}$
        \\ $k \gets k+1$
    }
}

\KwRet $\hat{x} = x_{prop}^{k}$
\end{algorithm}

\begin{figure}%
    \centering
    \subfloat[\footnotesize \centering Experimental scan positions (blue)]{{\includegraphics[width=3.4cm]{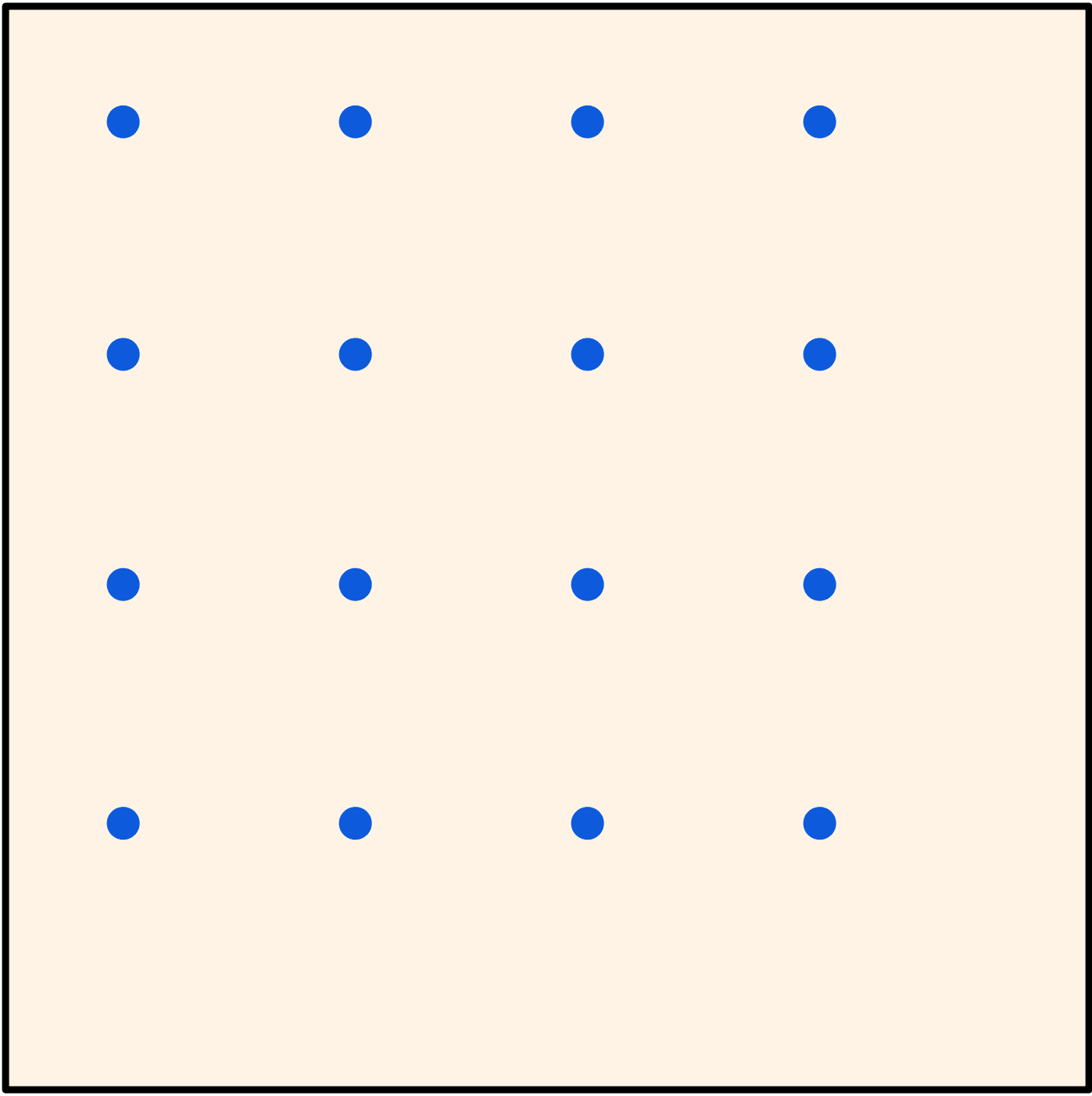} }}%
    \qquad
    \qquad
    \subfloat[\footnotesize \centering Oversampled simulated scan positions (red)]{{\includegraphics[width=3.4cm]{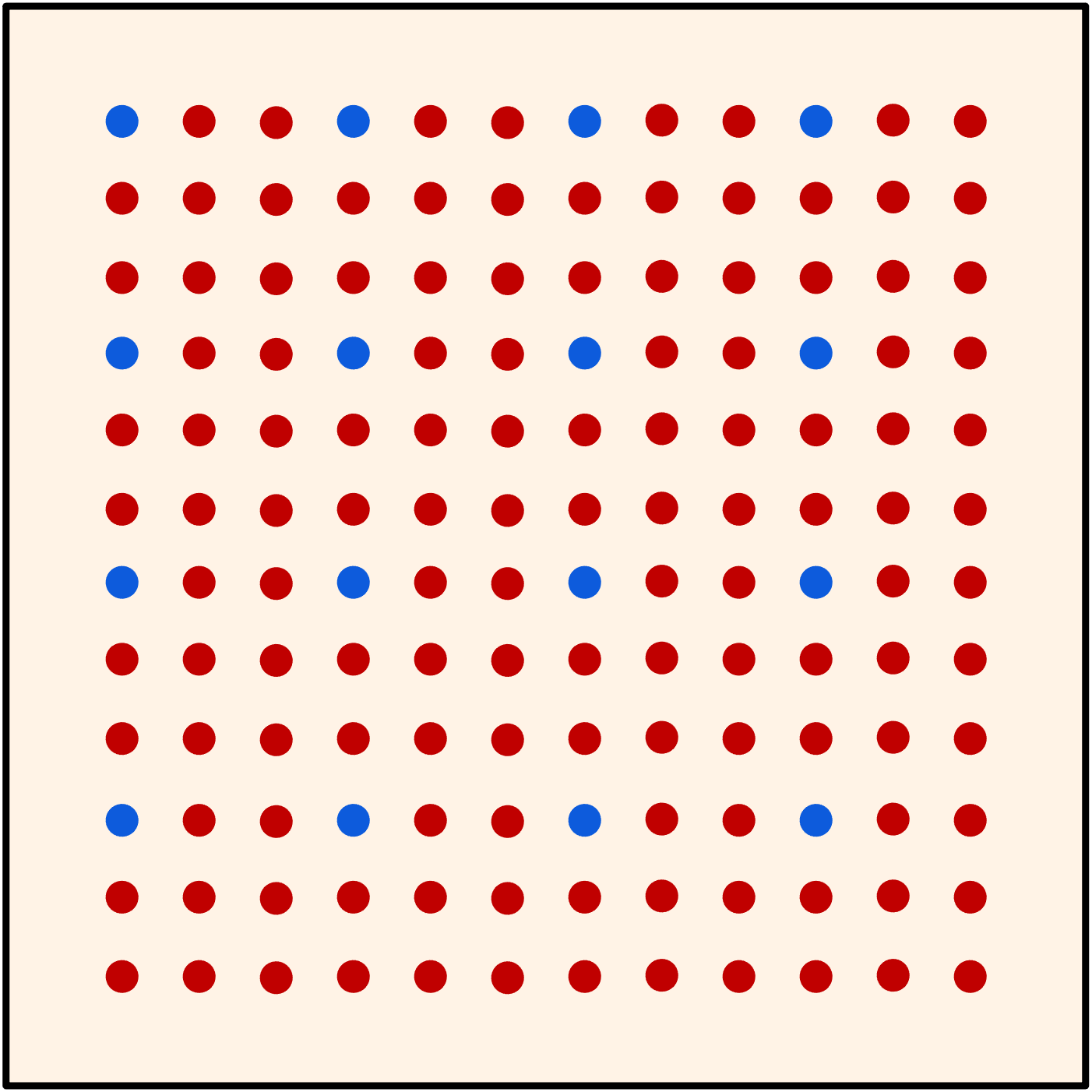} }}%
    \caption{Raster scan positions for the experimental data (left) and the simulated oversampled scan positions for the reconstruction taken from the output of the PtychoGAN algorithm (right). Scan positions for the real data are shown in blue, and the additional scan positions that are simulated are shown in red.}%
    \label{fig:scans}%
\end{figure}

ePIE is an algorithm producing high-quality reconstructions when enough overlap exists between scanning positions. The reconstructions have high resolution in the small details, avoiding possible minimization errors that deep learning approaches might suffer. On the other hand, the PtychoGAN approach in \cite{Barutcu22} is significantly more successful than ePIE in low or no overlap cases, but it might suffer distortions in details when there is enough overlap.

We propose an approach where we suggest combining the strengths of both methods. The proposed approach uses PtychoGAN for an initial reconstruction for insufficient overlap data. Then, using this reconstruction and the forward model of ptychography shown in Eqs~\ref{Eq:forward1} and \ref{Eq:forward2}, simulation data is acquired using a smaller step size; that is, a higher overlap scanning is simulated using the reconstruction. After the simulation, diffraction patterns corresponding to the original scan positions are replaced with the real data. Using this "oversampled" data, the ePIE algorithm is used to get reconstructions. This allows the ePIE to have enough overlap and increases the reconstruction quality compared to PtychoGAN results avoiding artifacts caused by the network.

If the diffraction patterns from low-overlap experimental data are called $d_{m, low}$, initial reconstruction $\hat{x}_{low}$ can be acquired using Eq.~\ref{Eq:converted}:
\begin{eqnarray}
    \label{Eq:prop1}
    \hat{z}, \hat{\theta} &=& \argmin_{z, \theta} \sum_m \| |F(P_ \odot e^{jG_{\theta}(z)}|^2 - 2d_{m, low} log|F(P_m \odot e^{jG_{\theta}(z)}|\|_1 + \lambda_1 \| \nabla (G_{\theta}(z)) \|_1 + \lambda_2 log (1 - D((G_{\theta}(z))
    \\ \hat{x}_{low} &=& e^{G_{\hat{\theta}}(\hat{z})}.
    \label{Eq:prop2}
\end{eqnarray}
where $m = 0, 1, 2, ... M$ denotes the positions for the low overlap experiments. Then using Eqs.~\ref{Eq:forward1} and \ref{Eq:forward2}, we can simulate the diffraction patterns from the reconstruction $\hat{x}_{low}$, with a higher overlap percentage such that the step size of the experiments is a multiple of the simulation step size which can be denoted by oversampling constant $Y$. Thus, if the positions for the high overlap simulations are denoted by $n = 0, 1, 2, ... N$, the relationship becomes $N = Y^2 M$. The process for the simulations can be shown as follows:
\begin{equation}
\label{Eq:prop3}
    d_{n, high} = |F (P_n \odot \hat{x}_{low})|^2.
\end{equation} 
The diffraction patterns acquired $d_{n, high}$ contain all simulated data. However, $M$ of these diffraction patterns have corresponding diffraction data from the experiments in $d_{m, low}$. Thus, we can replace these diffraction data and create the set $d_{n, mixed}$. Using this set of diffraction patterns, we can initialize $x_{prop}^{0}$ with an empty matrix and apply the ePIE reconstruction described in Eqs.~\ref{Eq:epie1}, \ref{Eq:epie2}, and \ref{Eq:epie3}:
\begin{equation}
\label{Eq:prop4}
    x_{prop}^{k+1} = x_{prop}^{k} + \alpha \frac{P^*_n}{|P_n|^2_{max}} \{(F^{-1} \{\sqrt{d_{n, mixed}} \frac{F P_n \odot x_{prop}^{k}}{|F P_n \odot x_{prop}^{k}|}\}) - (P_n \odot x_{prop}^{k})\}
\end{equation} 
While updating the object using Eq.~\ref{Eq:prop4}, the simulated and the experimental data are weighted differently, forcing the solution to converge to different minima. If the diffraction pattern corresponds to the experimental data for each $k^{th}$ position, the $\alpha$ value in Eq.~\ref{Eq:prop4} is set to 1. However, if the diffraction pattern corresponds to the oversampled simulation data, the $\alpha$ value is set to $\frac{1}{w}$. In that way, with the increased weight $w$, the solution converges to the pure ePIE solution, and with the decreased weight $w$, the solution converges to the PtychoGAN solution.

The proposed solution increases the reconstruction quality using the combination of ePIE and PtychoGAN with oversampling. The novel algorithm we propose in this paper is shown in Algorithm~\ref{alg:alg1} and can be summarized as:

\begin{itemize}
\item Get an initial reconstruction for M diffraction patterns using the PtychoGAN algorithm (Eqs.~\ref{Eq:prop1} and \ref{Eq:prop2}).
\item Apply the forward ptychography model on the reconstruction to get simulated data with $N = Y^2 M$ scan points (Eq.~\ref{Eq:prop3}).
\item Replace $M$ of the $N$ diffraction patterns with the experimental data for acquiring oversampled diffractions.
\item Apply the ePIE algorithm on the oversampled data for the final reconstruction (Eq.~\ref{Eq:prop4}).
\end{itemize}

Furthermore, we propose that the reconstruction quality further improved with increased fidelity to the data, which is achieved by repeating lines 2-11 with the reconstruction achieved in Algorithm~\ref{alg:alg1}. That is, after acquiring the reconstruction, it can be used as an initial point, and the forward model can be applied to repeat the ePIE reconstructions. This achieves higher fidelity to reconstruction data; however, the artifacts are also amplified with the increased number of iterations.

\section{Experiments and Results}

In this section, we explain our simulation setup for testing the proposed algorithm, show simulation results and how they vary with changing parameters, and compare the algorithm with previously proposed approaches under different conditions to show its effectiveness.

\subsection{Simulation Setup}

To test the effectiveness of the approach, we performed simulations on face images to be able to have a fair comparison with the architecture proposed in \cite{Barutcu22}. The structure for the PtychoGAN algorithm is followed as described, and the StylgaGAN2 network \cite{stylegan2} is trained on Flickr-Faces-HQ (FFHQ) \cite{stylegan} dataset for the Deep Generative Prior. We utilized the PyTorch implementation for the StyleGAN2 architecture \cite{Wang2020}, and we chose two different test images which do not have smooth backgrounds resulting in imperfect reconstructions by the PtychoGAN architecture.

For the ptychography simulations, we used a scaled version of the probe used in the PtychoGAN implementation with a diameter size $60$ \cite{Nashed_Peterka_Deng_Jacobsen_2017} pixels. The face images scanned are converted into gray-scale images and then converted to complex matrices with the size of $240\times 240$ pixels by setting the complex numbers' phase values to the gray-scale images and the amplitude values to a constant (1). The forward model described in Eqs.~\ref{Eq:forward1} and \ref{Eq:forward2} is applied to get the diffraction data with low or no overlap percentages. We added Poisson noise for testing the performance under noisy conditions. 

To compare the effectiveness of the algorithm different under different conditions, we used different levels of overlap percentages. $25\%$, $0\%$, $-25\%$, and $-50\%$ overlap percentages corresponding to probe step sizes of $45$, $60$, $75$, and $90$ are chosen, which either have a low overlap or no overlap at all. After the initial reconstruction using PtychoGAN, we used oversampling in different ratios ranging between 2 and 6 times the original sampling amount, and the oversampling ratio is shown by $Y$. Similarly, different values of weights ($w$) are used to show the effect of weighting.

\subsection{Reconstructions and Comparison}

\setlength\tabcolsep{3.2pt}
\begin{table}[!b]
\small
\caption{MSE values for reconstructions in Figs~\ref{fig:weights5} and \ref{fig:weights6} using the proposed algorithm with changing weights ($w$).}
\centering
\begin{tabular}{@{}x{1.5cm}|x{2cm}|x{1.5cm}|x{1cm}|x{1cm}|x{1cm}|x{1cm}|x{1cm}|x{1cm}|x{1cm}|x{1cm}|x{1cm}|@{}}
\multicolumn{1}{|c|}{\textbf{MSE ($x10^{-3}$)}}&
\multicolumn{1}{c|}{\multirow{2}{*}{\textbf{PtychoGAN}}} &
\multicolumn{9}{c|}{\textbf{Proposed Method}} &
\multicolumn{1}{c|}{\multirow{2}{*}{\textbf{ePIE}}} \\
\cmidrule{1-1} \cmidrule{3-11}
\multicolumn{1}{|c|}{\textbf{Object}} &
 &
\multicolumn{1}{c|}{{$w=1e-4$}} &
\multicolumn{1}{c|}{{$w=0.1$}} &
\multicolumn{1}{c|}{{$w=1$}} &
\multicolumn{1}{c|}{{$w=2$}} &
\multicolumn{1}{c|}{{$w=5$}} &
\multicolumn{1}{c|}{{$w=10$}} &
\multicolumn{1}{c|}{{$w=20$}} &
\multicolumn{1}{c|}{{$w=100$}} &
\multicolumn{1}{c|}{{$w=1e4$}} &
 \\
% \cmidrule(l){2-9} 
\midrule
% \multicolumn{1}{|c|}{Method} & {ePIE} & {Prop} & {ePIE} & {Prop} \\ \midrule
\multicolumn{1}{|c|}{{Image 1}} &
6.32 & 5.75 & 5.60 & 4.51 & 4.15 & 4.60 & 5.20 & \textbf{3.04} & 6.94 & 66.96 & 63.09\\
\multicolumn{1}{|c|}{{Image 2}} &
2.31 & 2.43 & 2.30 & 1.87 & 1.53 & 1.22 & 1.07 & \textbf{0.86} & 53.17 & 198.15 & 177.69\\
\midrule
\end{tabular}
\label{tab:weights_tb}
\end{table}

% \setlength\tabcolsep{3.2pt}
% \begin{table}[hb]
% \small
% \caption{Caption}
% \centering
% \begin{tabular}{@{}c|x{2cm}|x{1.5cm}|x{1cm}|x{1cm}|x{1cm}|x{1cm}|x{1cm}|x{1cm}|x{1cm}|x{1cm}|x{1cm}|@{}}
% \multicolumn{1}{|c|}{\textbf{SSIM}}&
% \multicolumn{1}{c|}{\multirow{2}{*}{\textbf{PtychoGAN}}} &
% \multicolumn{9}{c|}{\textbf{Proposed Method}} &
% \multicolumn{1}{c|}{\multirow{2}{*}{\textbf{ePIE}}} \\
% \cmidrule{1-1} \cmidrule{3-11}
% \multicolumn{1}{|c|}{\textbf{Object}} &
%  &
% \multicolumn{1}{c|}{{$w=1e-4$}} &
% \multicolumn{1}{c|}{{$w=0.1$}} &
% \multicolumn{1}{c|}{{$w=1$}} &
% \multicolumn{1}{c|}{{$w=2$}} &
% \multicolumn{1}{c|}{{$w=5$}} &
% \multicolumn{1}{c|}{{$w=10$}} &
% \multicolumn{1}{c|}{{$w=20$}} &
% \multicolumn{1}{c|}{{$w=100$}} &
% \multicolumn{1}{c|}{{$w=1e4$}} &
%  \\
% % \cmidrule(l){2-9} 
% \midrule
% % \multicolumn{1}{|c|}{Method} & {ePIE} & {Prop} & {ePIE} & {Prop} \\ \midrule
% \multicolumn{1}{|c|}{{Image 1}} &
% 0.863 & 0.860 & 0.863 & 0.880 & 0.892 & 0.906 & 0.914 & \textbf{0.926} & 0.788 & 0.254 & 0.256\\
% \multicolumn{1}{|c|}{{Image 2}} &
% 0.919 & 0.918 & 0.920 & 0.930 & 0.940 & 0.953 & \textbf{0.960} & 0.959 & 0.584 & 0.155 & 0.173\\
% \hline
% \end{tabular}
% \label{tab:weights_tb}
% \end{table}

We test the robustness and effectiveness of the process by changing parameters and showing their effects, applying different overlap percentages, observing under noisy conditions, and comparing with the state-of-the-art approaches. All reconstructions shown for PtychoGAN and the initial reconstructions for the proposed approach follow the methodology described in \cite{Barutcu22}. For the ePIE updates after the initial reconstruction, the algorithm is continued until convergence, usually between 2000-6000 ePIE iterations. 

\begin{figure*}[!t]
    \centering
    \setlength{\tabcolsep}{\szb}

    \begin{tabular}{cccccc}
    {PtychoGAN} & {$w=0.0001$} & {$w=0.1$} &  {$w=1$} & {$w=2$} & {$w=5$}
    \\
    \includegraphics[width=\sza\columnwidth/2]{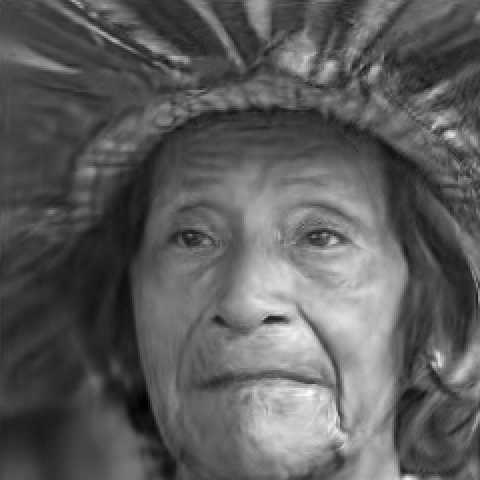} &   
    \includegraphics[width=\sza\columnwidth/2]{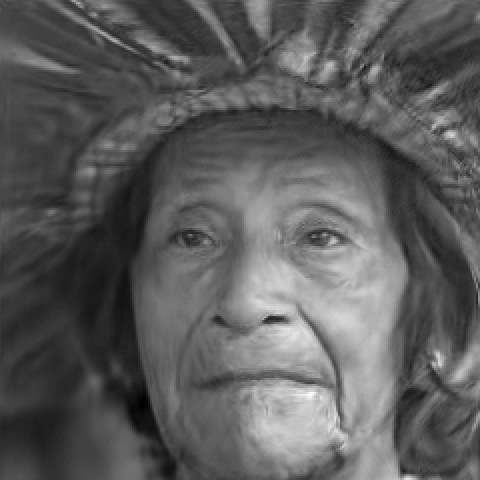} &   
    \includegraphics[width=\sza\columnwidth/2]{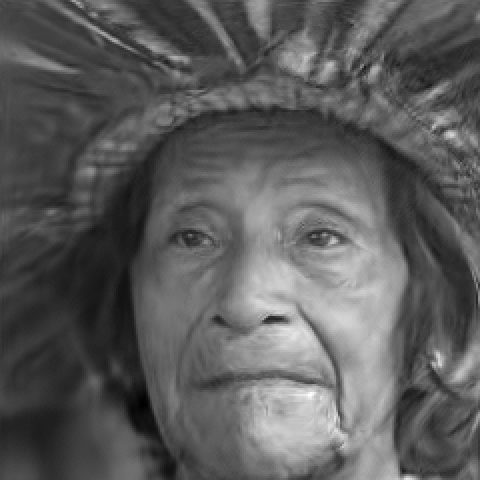} &   
    \includegraphics[width=\sza\columnwidth/2]{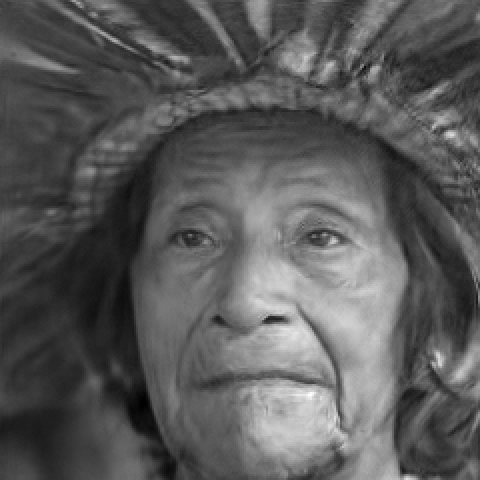} &   
    \includegraphics[width=\sza\columnwidth/2]{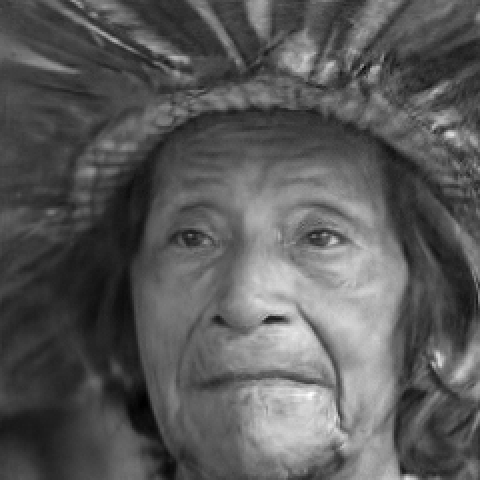} &
    \includegraphics[width=\sza\columnwidth/2]{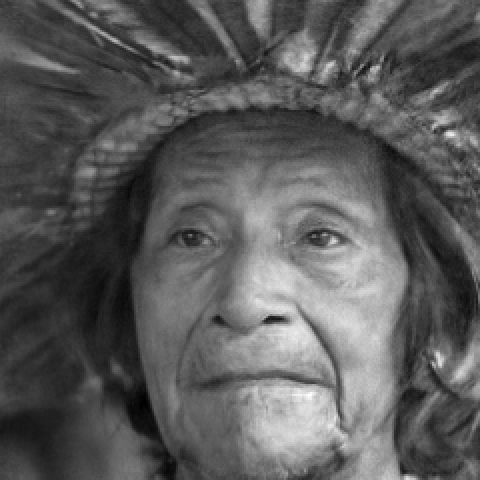}
    \\
    {$w=10$} & {$w=20$} &  {$w=100$} & {$w=10000$} & {ePIE} & {GT}
    \\
    \includegraphics[width=\sza\columnwidth/2]{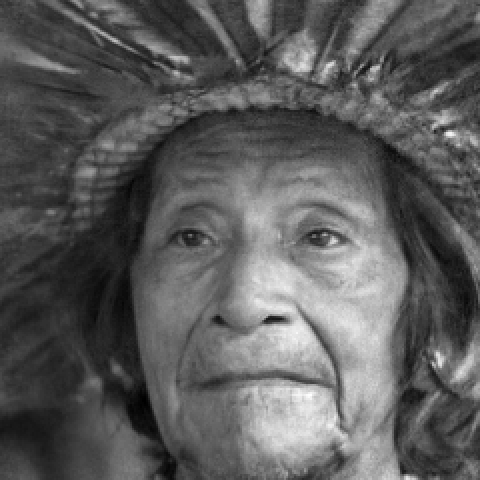} &   
    \includegraphics[width=\sza\columnwidth/2]{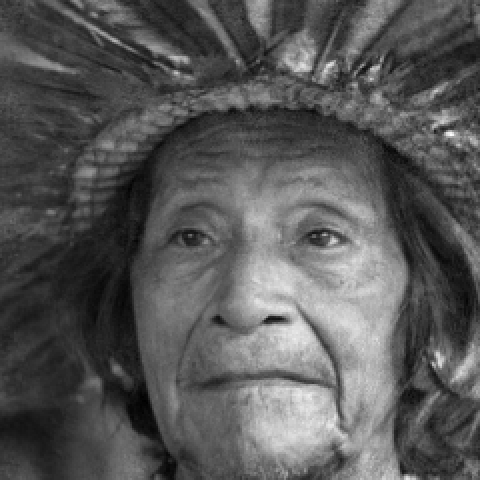} &   
    \includegraphics[width=\sza\columnwidth/2]{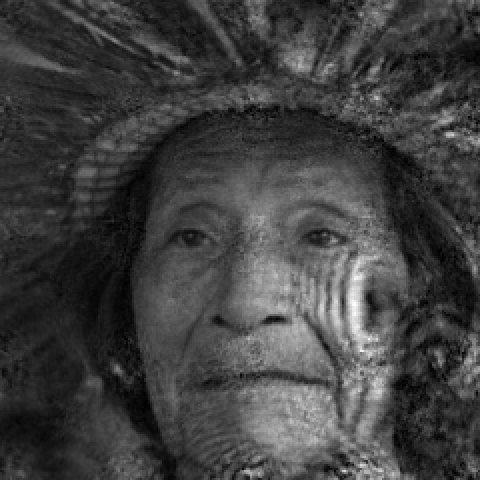} &   
    \includegraphics[width=\sza\columnwidth/2]{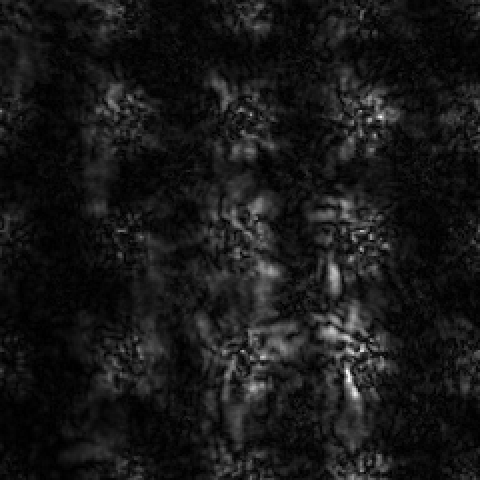} &  
    \includegraphics[width=\sza\columnwidth/2]{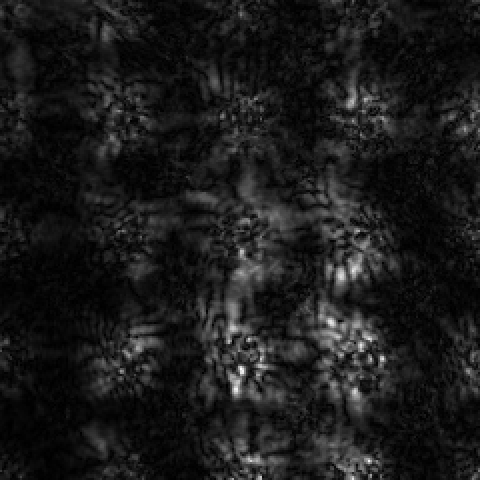} &  
    \includegraphics[width=\sza\columnwidth/2]{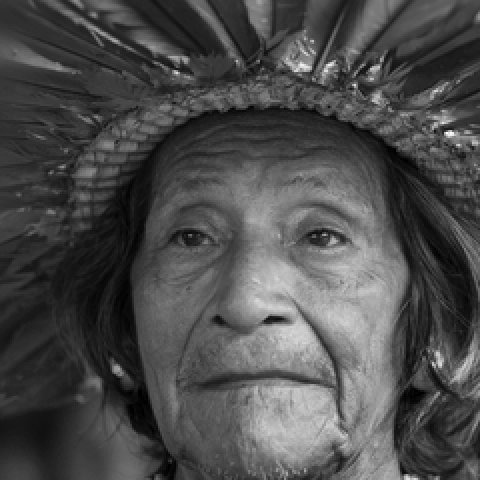}
    \vspace{-10pt}
\end{tabular}
\caption{Reconstructions using the proposed algorithm with changing weights ($w$) that controls the update amount derived from the simulated data and the real data during ePIE reconstruction. Higher $w$ means that algorithm is more dependent on the experimental data and it omits the simulated data. Here, $0\%$ overlap is used with 3 times over-sampling ($Y=3$), and the results are compared with PtychoGAN and ePIE algorithms, and ground truth (GT) for Image 1.}
 \label{fig:weights5}
\end{figure*}

\begin{figure*}[t]
    \centering
    \setlength{\tabcolsep}{\szb}

    \begin{tabular}{cccccc}
    {PtychoGAN} & {$w=0.0001$} & {$w=0.1$} &  {$w=1$} & {$w=2$} & {$w=5$}
    \\
    \includegraphics[width=\sza\columnwidth/2]{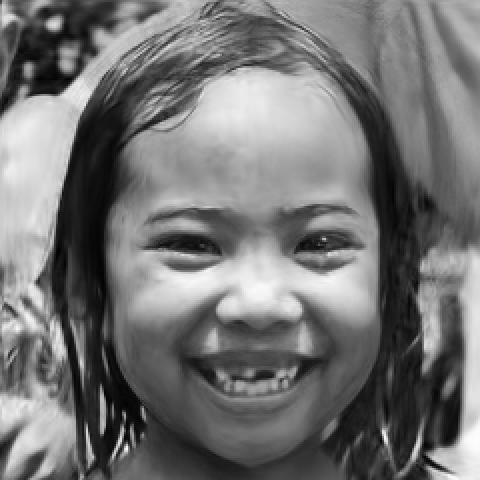} &   
    \includegraphics[width=\sza\columnwidth/2]{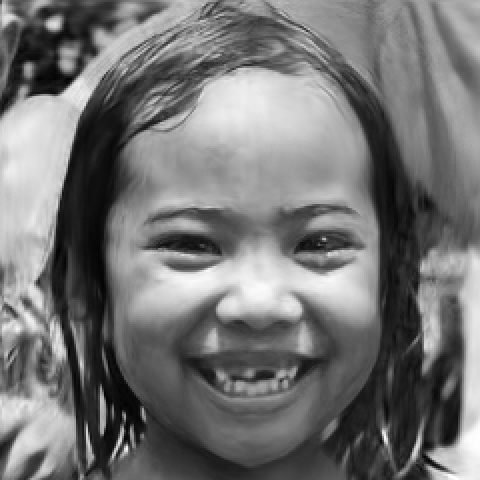} &  
    \includegraphics[width=\sza\columnwidth/2]{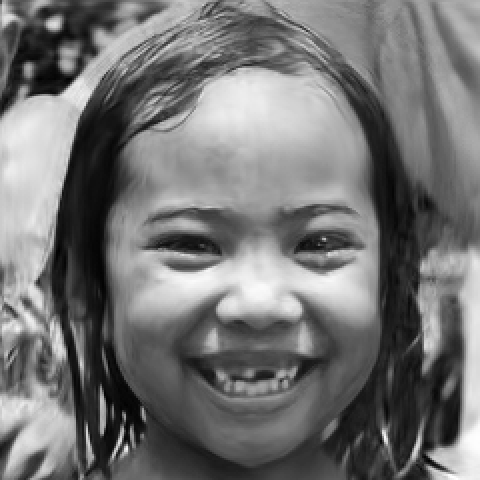} &  
    \includegraphics[width=\sza\columnwidth/2]{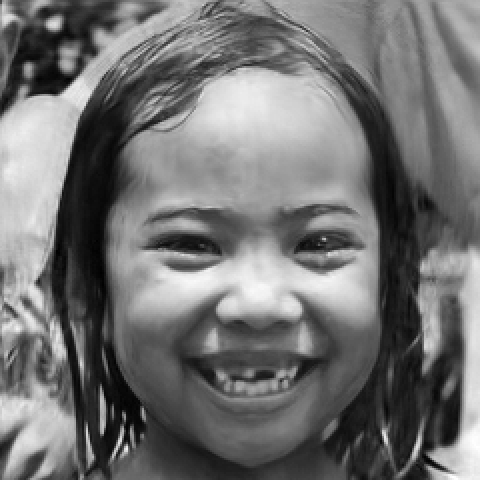} &   
    \includegraphics[width=\sza\columnwidth/2]{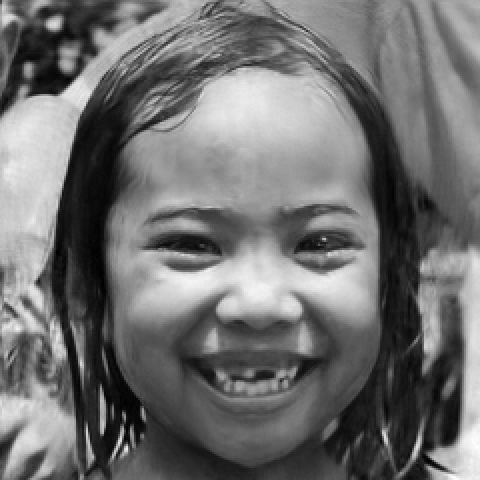} &  
    \includegraphics[width=\sza\columnwidth/2]{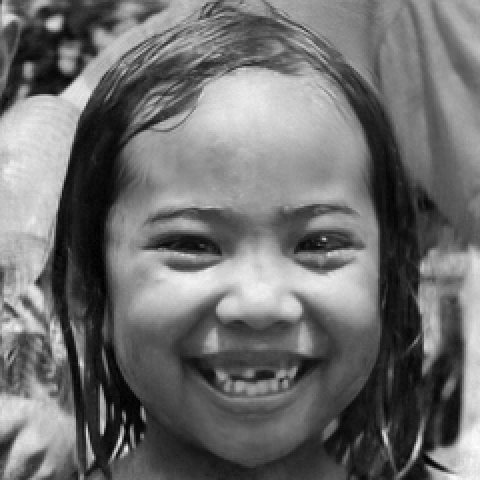}
    \\
    {$w=10$} & {$w=20$} &  {$w=100$} & {$w=10000$} & {ePIE} & {GT}
    \\
    \includegraphics[width=\sza\columnwidth/2]{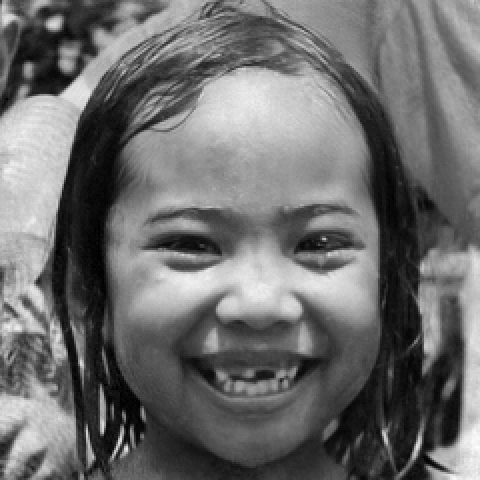} &   
    \includegraphics[width=\sza\columnwidth/2]{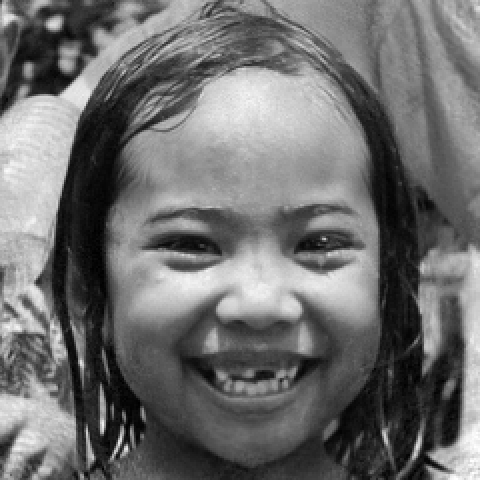} &  
    \includegraphics[width=\sza\columnwidth/2]{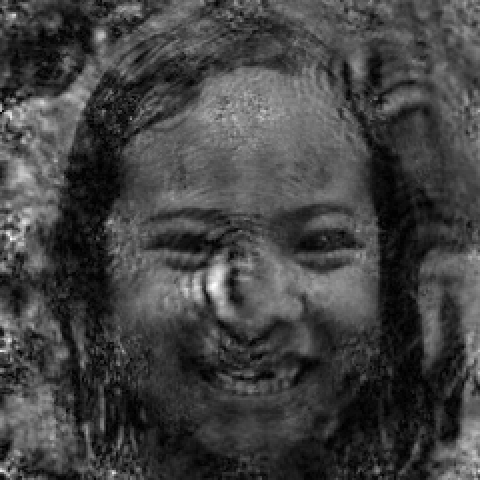} &   
    \includegraphics[width=\sza\columnwidth/2]{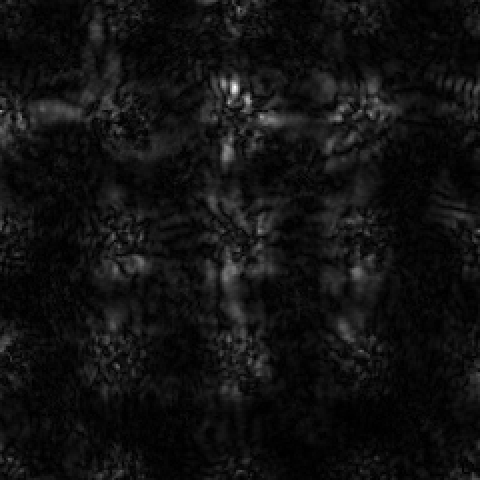} &   
    \includegraphics[width=\sza\columnwidth/2]{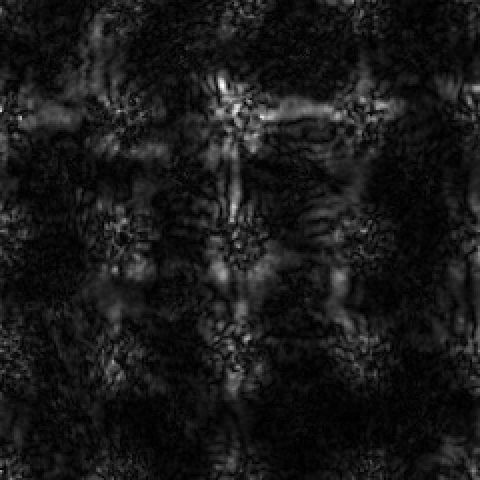} &  
    \includegraphics[width=\sza\columnwidth/2]{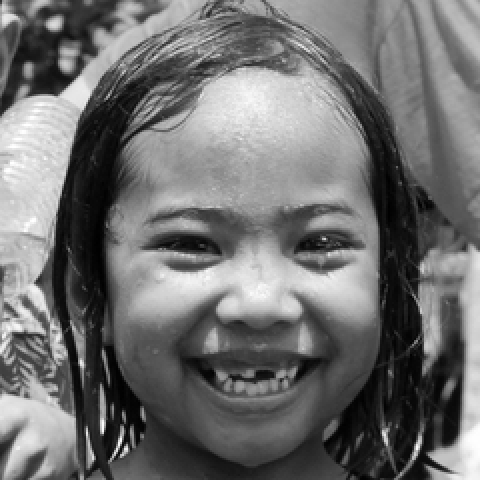}
    \vspace{-10pt}
\end{tabular}
\caption{Reconstructions using the proposed algorithm with changing weights ($w$) and comparisons with PtychoGAN and ePIE algorithms for Image 2 using same parameters in Fig.~\ref{fig:weights5}}
 \label{fig:weights6}
\end{figure*}

In Figs.~\ref{fig:weights5} and \ref{fig:weights6}, we show the reconstructions with different weights $w$ from diffraction data acquired by $0\%$ overlap. For the proposed approach, the oversampling is kept at $Y=3$; that is, after the initial reconstruction by PtychoGAN with a step size of $60$ pixels, artificial diffraction data is created using $20$ pixels of step size. This results in an oversampling ratio of $Y=3$ in one direction, resulting in $Y^2 = 9$ times more data for the following ePIE steps. We can see that the reconstruction quality increases compared to PtychoGAN, and ePIE fails to recover the object with no overlap.

The effects of the weights can be observed in Figs.~\ref{fig:weights5} and \ref{fig:weights6}. We can see a smooth transition from the PtychoGAN reconstruction to ePIE reconstruction with the increased weights. In very low values of $w$, i.e., 0.0001, we can see that the reconstruction is quite close to the PtychoGAN results. Similarly, for very high values of $w$, i.e., 100000, the reconstruction resembles the one of ePIE. Since these two approaches have different weaknesses and strengths, reconstructions with a $w$ value in between show better reconstructions, and the reconstruction quality can be increased by adjusting $w$. 

Table~\ref{tab:weights_tb} shows the mean-squared error (MSE) values of the reconstructions compared to the ground truth (GT) images. As can be observed, the optimal solution is given by $w=20$, which shows that if we give more weight to the experimental data instead of the simulated data, we can get higher quality reconstructions.

\setlength\tabcolsep{3.2pt}
\begin{table}[!b]
\small
\caption{MSE values for reconstructions in Fig~\ref{fig:ratios} using the proposed algorithm with changing over-sampling ratios ($Y$).}
\centering
\begin{tabular}{@{}c|x{2cm}|x{1cm}|x{1cm}|x{1cm}|x{1cm}|x{1cm}|@{}}
\multicolumn{1}{|c|}{\textbf{MSE ($x10^{-3}$)}}&
\multicolumn{1}{c|}{\multirow{2}{*}{\textbf{PtychoGAN}}} &
\multicolumn{4}{c|}{\textbf{Proposed Method}} &
\multicolumn{1}{c|}{\multirow{2}{*}{\textbf{ePIE}}} \\
\cmidrule{1-1} \cmidrule{3-6}
\multicolumn{1}{|c|}{\textbf{Object}} &
 &
\multicolumn{1}{c|}{{$Y=2$}} &
\multicolumn{1}{c|}{{$Y=3$}} &
\multicolumn{1}{c|}{{$Y=4$}} &
\multicolumn{1}{c|}{{$Y=6$}} &
 \\
% \cmidrule(l){2-9} 
\midrule
% \multicolumn{1}{|c|}{Method} & {ePIE} & {Prop} & {ePIE} & {Prop} \\ \midrule
\multicolumn{1}{|c|}{{Image 1}} &
6.32 & \textbf{2.32} & 3.04 & 4.35 & 3.84 & 63.08\\
\multicolumn{1}{|c|}{{Image 2}} &
2.31 & 7.98 & \textbf{1.04} & 1.15 & 1.26 & 177.69\\
\midrule
\end{tabular}
\label{tab:ratios_tb}
\end{table}

% \setlength\tabcolsep{3.2pt}
% \begin{table}[hb]
% \small
% \caption{Caption}
% \centering
% \begin{tabular}{@{}c|x{2cm}|x{1cm}|x{1cm}|x{1cm}|x{1cm}|x{1cm}|@{}}
% \multicolumn{1}{|c|}{\textbf{SSIM}} &
% \multicolumn{1}{c|}{\multirow{2}{*}{\textbf{PtychoGAN}}} &
% \multicolumn{4}{c|}{\textbf{Proposed Method}} &
% \multicolumn{1}{c|}{\multirow{2}{*}{\textbf{ePIE}}} \\
% \cmidrule{1-1} \cmidrule{3-7}
% \multicolumn{1}{|c|}{\textbf{Object}} &
%  &
% \multicolumn{1}{c|}{{$Y=2$}} &
% \multicolumn{1}{c|}{{$Y=3$}} &
% \multicolumn{1}{c|}{{$Y=4$}} &
% \multicolumn{1}{c|}{{$Y=6$}} &
%  \\
% % \cmidrule(l){2-9} 
% \midrule
% % \multicolumn{1}{|c|}{Method} & {ePIE} & {Prop} & {ePIE} & {Prop} \\ \midrule
% \multicolumn{1}{|c|}{{Image 1}} &
% 0.863 & 0. & 0. & 0. & 0. & 0.256\\
% \multicolumn{1}{|c|}{{Image 2}} &
% 0.919 & 0. & 0. & 0. & 0. & 0.173\\
% \hline
% \end{tabular}
% \label{tab:ratios_tb}
% \end{table}

In addition to the effect of the weighting, we tested the effect of the oversampling ratio. Although the effect of the oversampling ratio is similar to the effect of weighting, it is not the same. Too much oversampling will result in an increased number of diffraction patterns from the simulated data, and we expect the reconstruction to look similar to the PtychoGAN reconstruction with the increased oversampling. Similarly, it will diverge from the PtychoGAN reconstruction and get closer to ePIE reconstruction when the oversampling ratio is small. This is similar to what we observe in weighting. However, the shift will not be as drastic since the effect of the experimental data is mostly independent of the update of the simulated data. 

\begin{figure*}[!t]
    \centering
    \setlength{\tabcolsep}{\szb}

    \begin{tabular}{ccccccc}
    {PtychoGAN} & {$Y=2$} & {$Y=3$} &  {$Y=4$} & {$Y=6$} & {ePIE} & {GT}
    \\
    \includegraphics[width=\sza\columnwidth/2]{figs/pg/im5/discLoss0_realProbe_noisePOISSON0_tvnewmult0_probe120_step60.png} &    
    \includegraphics[width=\sza\columnwidth/2]{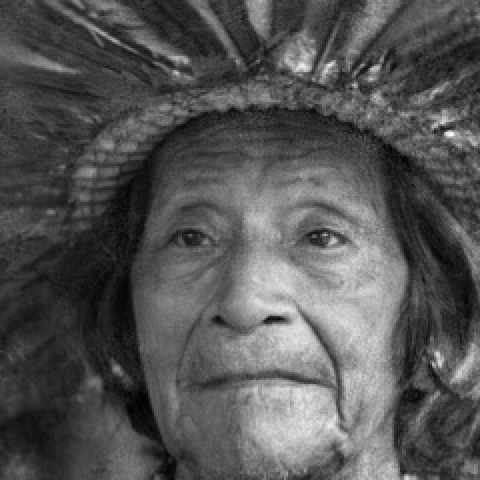} &   
    \includegraphics[width=\sza\columnwidth/2]{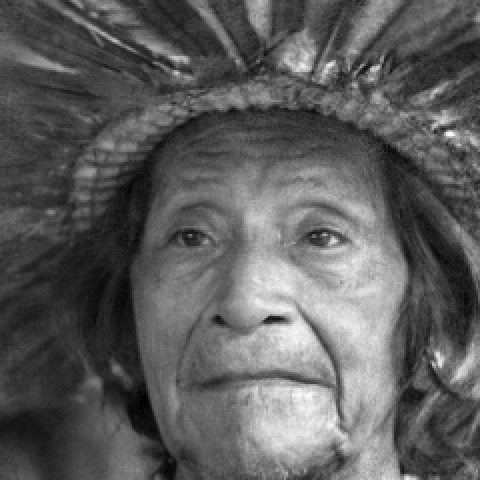} &   
    \includegraphics[width=\sza\columnwidth/2]{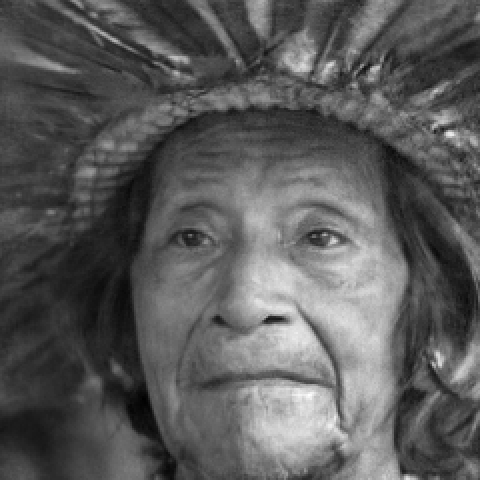} &   
    \includegraphics[width=\sza\columnwidth/2]{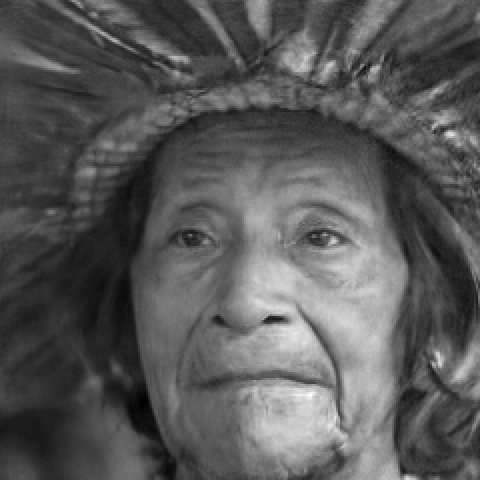} &
    \includegraphics[width=\sza\columnwidth/2]{figs/epie/im5/initial60.0to60_noRandShift_l1Loss_weight10_combinedmultiple1.0_discLoss0_realProbe_noisePOISSON0_tvnewmult0_probe120_step60.png} &  
    \includegraphics[width=\sza\columnwidth/2]{figs/gt5.png}
    \\
    \includegraphics[width=\sza\columnwidth/2]{figs/pg/im6/discLoss0_realProbe_noisePOISSON0_tvnewmult0_probe120_step60.png} &  
    \includegraphics[width=\sza\columnwidth/2]{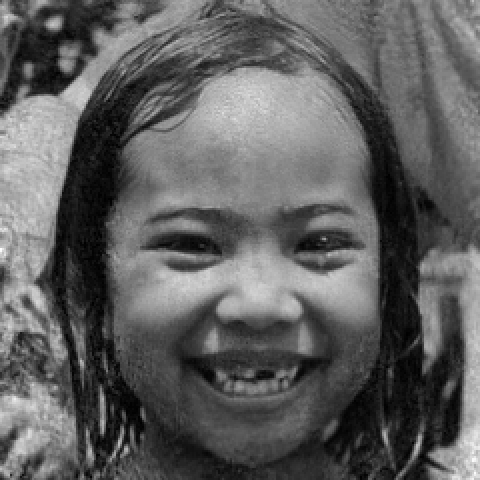} &    
    \includegraphics[width=\sza\columnwidth/2]{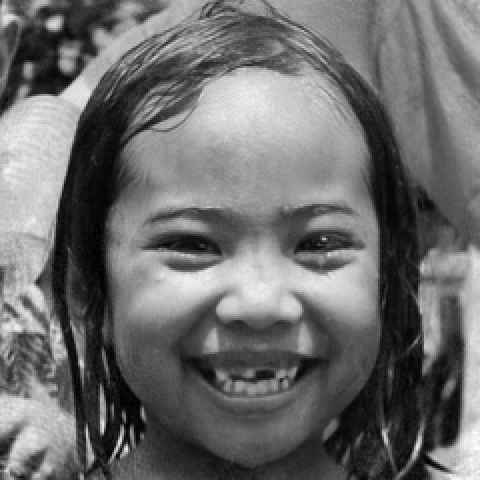} &   
    \includegraphics[width=\sza\columnwidth/2]{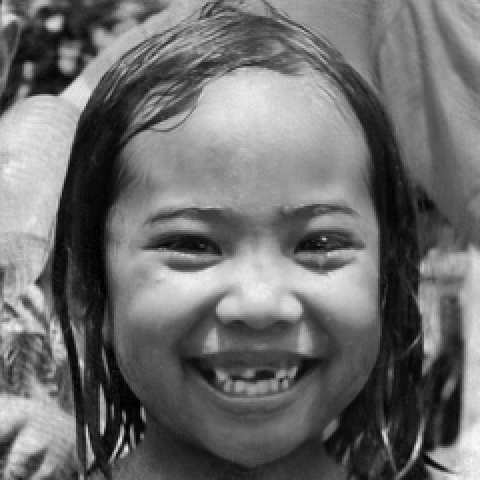} &   
    \includegraphics[width=\sza\columnwidth/2]{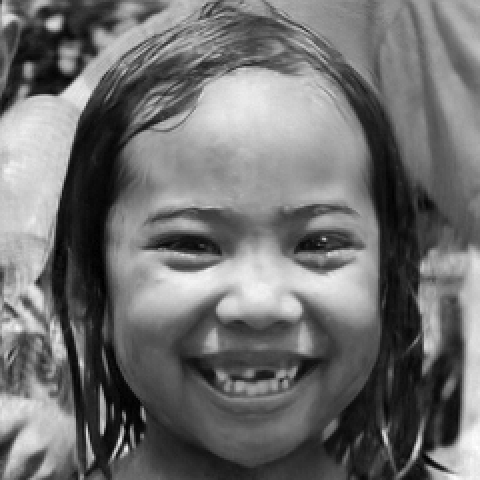} &   
    \includegraphics[width=\sza\columnwidth/2]{figs/epie/im6/initial60.0to60_noRandShift_l1Loss_weight10_combinedmultiple1.0_discLoss0_realProbe_noisePOISSON0_tvnewmult0_probe120_step60.png} &  
    \includegraphics[width=\sza\columnwidth/2]{figs/gt6.png}
    \vspace{-10pt}
\end{tabular}
\caption{Reconstructions using the proposed algorithm with changing over-sampling ratios ($Y$) that controls the amount of simulated data generated from PtychoGAN reconstructions Here, $0\%$ overlap is used with $w = 20$, and the results are compared with PtychoGAN and ePIE algorithms.}
 \label{fig:ratios}
\end{figure*}

The reconstructions for different oversampling ratios can be seen in Fig.~\ref{fig:ratios} together with corresponding reconstructions from other methods, and corresponding MSE values for the reconstructions can be observed in Table~\ref{tab:ratios_tb}. Although there are differences in reconstructions, the change is not too significant most of the time. However, we observe that, on average, $Y=3$ is a good value, and we used it for most of our reconstructions. An important effect here is the effect of the oversampling ratio on the reconstruction time. Increased oversampling affects the reconstruction time for ePIE updates directly, and this increase is proportional to the square of the increase in the oversampling ratio. Theoretically, we can simulate the diffraction patterns with huge overlap percentages; however, it is practically not feasible. The reconstructions seem to diverge from the ideal when the oversampling ratio goes above $Y=3$ from the results in Table~\ref{tab:ratios_tb}.

ePIE requires a significant amount of overlap for a successful reconstruction from x-ray ptychography data. When the overlap percentage is low such as $25\%$, or when there is no overlap between the adjacent scans, such as $0\%$, $-25\%$, and $-50\%$ overlaps, the reconstructions fail to show the structure of the object. PtychoGAN solves this issue by utilizing Deep Generative and Deep Image Priors (DGPs and DIPs). However, due to the generalizing nature of DGPs and due to the highly non-convex nature of DIPs, the reconstructions show some artifacts. These can be small distortions or introduced blur. 

The proposed approach utilizes the advantages of both methods to increase the reconstruction quality, and it is successful at it, which can be seen in Fig.~\ref{fig:compare_nf}. Here, we can see the reconstructions from ePIE, PtychoGAN, and the proposed approach together with the ground truth images for different overlap percentages. In these experiments, the oversampling ratio is kept at $Y=3$, and the weights are chosen to be $w=20$. Looking at the first columns, we see that ePIE fails to reconstruct the face completely for no overlap cases and shows significant distortions for the low overlap case. 

\setlength\tabcolsep{3.2pt}
\begin{table}[!b]
\footnotesize
\caption{MSE values for reconstructions in Fig~\ref{fig:compare_nf} using different methods with varying overlap ratios.}
\centering
\begin{tabular}{@{}c|x{2cm}|x{2cm}|x{2cm}||x{2cm}|x{2cm}|x{2cm}|@{}}
\multicolumn{1}{|c|}{\textbf{MSE ($x10^{-3}$)}}&
\multicolumn{3}{c||}{\textbf{Image 1}} & \multicolumn{3}{c|}{\textbf{Image 2}} \\
% \cmidrule(l){2-9} 
\midrule
\multicolumn{1}{|c|}{Method} & {ePIE} & {PtychoGAN} & {Proposed}  & {ePIE} & {PtychoGAN} & {Proposed} \\ \midrule
\multicolumn{1}{|c|}{Overlap = $25\%$} & 2.68 & 6.77 & \textbf{2.18} & 43.71 & 4.58 & \textbf{0.78} \\
\multicolumn{1}{|c|}{Overlap = $0\%$} & 63.09 & 6.32 & \textbf{3.04} & 177.69 & 2.31 & \textbf{1.04} \\
\multicolumn{1}{|c|}{Overlap = $-25\%$} & 97.45 & 6.83 & \textbf{2.70} & 232.06 & 3.56 & \textbf{2.77} \\
\multicolumn{1}{|c|}{Overlap = $-50\%$} & 95.69 & 8.32 & \textbf{2.89} & 227.65 & 5.94 & \textbf{3.86} \\
\midrule
\end{tabular}
\label{tab:compare_nf_tb}
\end{table}

% \setlength\tabcolsep{3.2pt}
% \begin{table}[!b]
% \footnotesize
% \caption{Caption}
% \centering
% \begin{tabular}{@{}c|x{2cm}|x{2cm}|x{2cm}||x{2cm}|x{2cm}|x{2cm}|@{}}
% \multicolumn{1}{|c|}{\textbf{SSIM}} &
% \multicolumn{3}{c||}{\textbf{Image 1}} & \multicolumn{3}{c|}{\textbf{Image 2}} \\
% % \cmidrule(l){2-9} 
% \midrule
% \multicolumn{1}{|c|}{Method} & {ePIE} & {PtychoGAN} & {Proposed}  & {ePIE} & {PtychoGAN} & {Proposed} \\ \midrule
% \multicolumn{1}{|c|}{Overlap = $25\%$} & 0.861 & 0.878 & \textbf{0.961} & 0.668 & 0.898 & \textbf{0.} \\
% \multicolumn{1}{|c|}{Overlap = $0\%$} & 0.256 & 0.863 & \textbf{0.913} & 0.173 & 0.919 & \textbf{0.} \\
% \multicolumn{1}{|c|}{Overlap = $-25\%$} & 0.144 & 0.753 & \textbf{0.876} & 0.080 & 0.885 & \textbf{0.} \\
% \multicolumn{1}{|c|}{Overlap = $-50\%$} & 0.154 & 0.831 & \textbf{0.846} & 0.101 & 0.827 & \textbf{0.} \\
% \hline
% \end{tabular}
% \label{tab:compare_nf_tb}
% \end{table}

PtychoGAN does significantly better in many cases. However, it is not free of issues. First, although it shows significant improvement ins the face area, the background is hard for the algorithm to reconstruct completely since it lacks the DGPs in these areas. Then, it suffers from distortions in some cases, most apparent in the $50\%$ overlap cases, and loses the structural integrity of the objects. Also, since it is hard for the algorithm not to be stuck at a local minimum during convergence, it might result in half-successful reconstructions, which can be observed for the $25\%$ overlap case for image 2. However, the proposed approach increases the reconstruction quality by combining the approach with ePIE steps. As can be seen in the details of the reconstruction, overall sharpness increases leading to a lower error in the reconstruction. More importantly, if there is a distortion in the PtychoGAN reconstruction, the algorithm forces the image to remove the distortion and bring the structure back in the picture. This is most apparent in $-50\%$ overlap cases for both images where the PtychoGAN reconstruction shows significant distortions, but the proposed approach reconstructs the face characteristics in those distorted areas. The MSE values for these reconstructions can be seen in Table~\ref{tab:compare_nf_tb}. Similar to the qualitative examination results, quantitative results show a similar trend, and the proposed approach performs better than the other approaches in all cases. While in some cases, the improvement is small, in some cases, there is a significant increase in the reconstruction quality. Also, even though the increase in quality is inconsistent in PtychoGAN, we can see from the MSE values that the quality increase is more stable in the proposed approach.

\begin{figure*}[!t]
    \centering
    \setlength{\tabcolsep}{\szb}

    \begin{tabular}{ccccccccc}
    & {ePIE} & {PtychoGAN} & {Proposed} & {GT} 
    & {ePIE} & {PtychoGAN} & {Proposed} & {GT} 
    \vspace{2pt}
    \\
    \rotatebox{90}{\hspace{20pt}\emph{$25\%$}}
    \hspace{1pt} &
    \includegraphics[width=\sza\columnwidth/2]{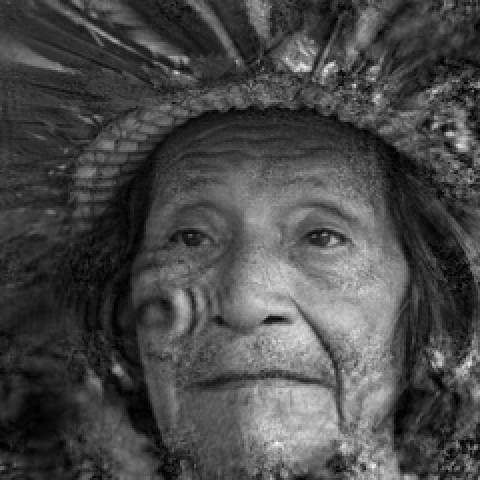} & 
    \includegraphics[width=\sza\columnwidth/2]{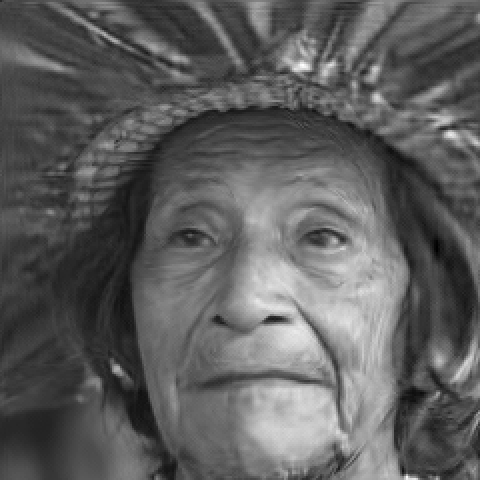} & 
    \includegraphics[width=\sza\columnwidth/2]{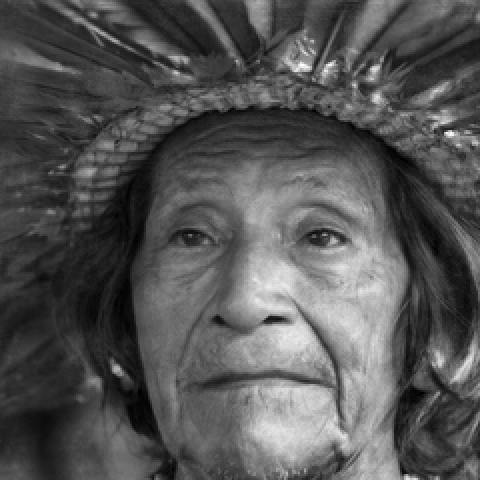} &
    \includegraphics[width=\sza\columnwidth/2]{figs/gt5.png} &
    \hspace{2pt}
    \includegraphics[width=\sza\columnwidth/2]{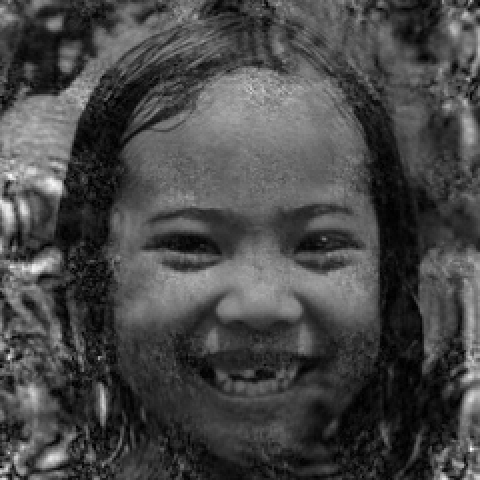} & 
    \includegraphics[width=\sza\columnwidth/2]{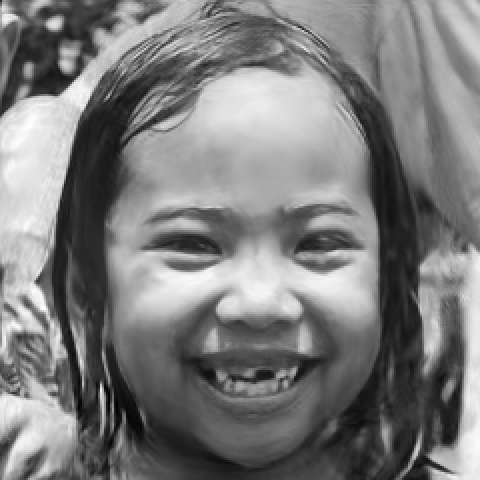} & 
    \includegraphics[width=\sza\columnwidth/2]{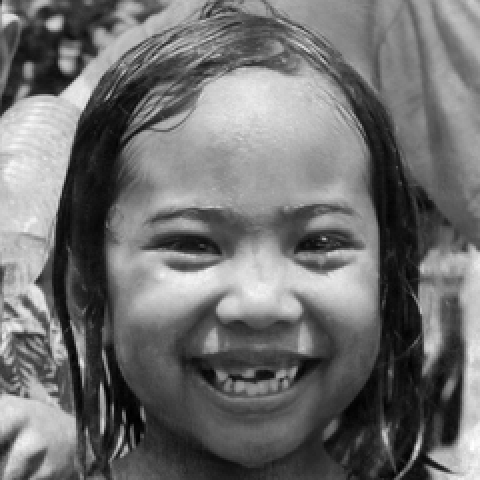} &
    \includegraphics[width=\sza\columnwidth/2]{figs/gt6.png}
    \\
    \rotatebox{90}{\hspace{24pt}\emph{$0\%$}}
    \hspace{1pt} &
    \includegraphics[width=\sza\columnwidth/2]{figs/epie/im5/initial60.0to60_noRandShift_l1Loss_weight10_combinedmultiple1.0_discLoss0_realProbe_noisePOISSON0_tvnewmult0_probe120_step60.png} & 
    \includegraphics[width=\sza\columnwidth/2]{figs/pg/im5/discLoss0_realProbe_noisePOISSON0_tvnewmult0_probe120_step60.png} & 
    \includegraphics[width=\sza\columnwidth/2]{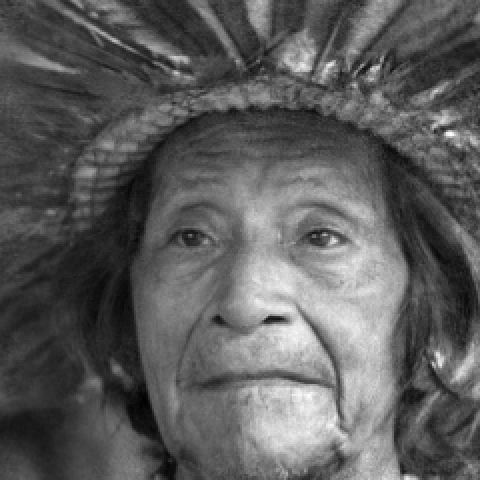} &
    \includegraphics[width=\sza\columnwidth/2]{figs/gt5.png} &
    \hspace{2pt}
    \includegraphics[width=\sza\columnwidth/2]{figs/epie/im6/initial60.0to60_noRandShift_l1Loss_weight10_combinedmultiple1.0_discLoss0_realProbe_noisePOISSON0_tvnewmult0_probe120_step60.png} & 
    \includegraphics[width=\sza\columnwidth/2]{figs/pg/im6/discLoss0_realProbe_noisePOISSON0_tvnewmult0_probe120_step60.png} & 
    \includegraphics[width=\sza\columnwidth/2]{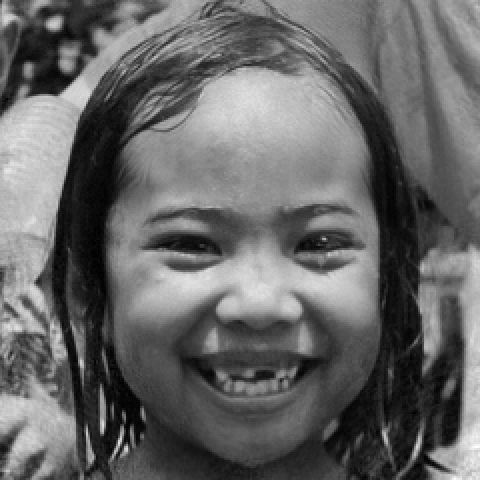} &
    \includegraphics[width=\sza\columnwidth/2]{figs/gt6.png}
    \\
    \rotatebox{90}{\hspace{14pt}\emph{$-25\%$}}
    \hspace{1pt} &
    \includegraphics[width=\sza\columnwidth/2]{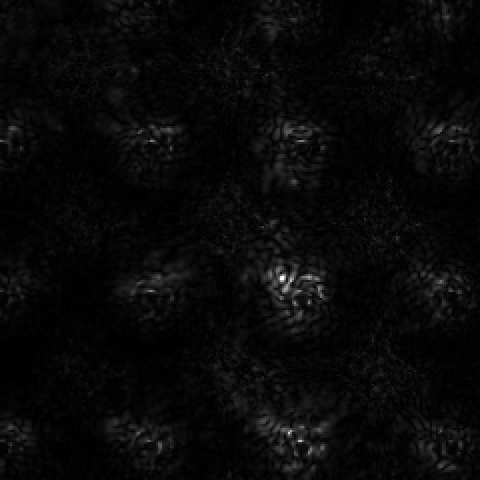} & 
    \includegraphics[width=\sza\columnwidth/2]{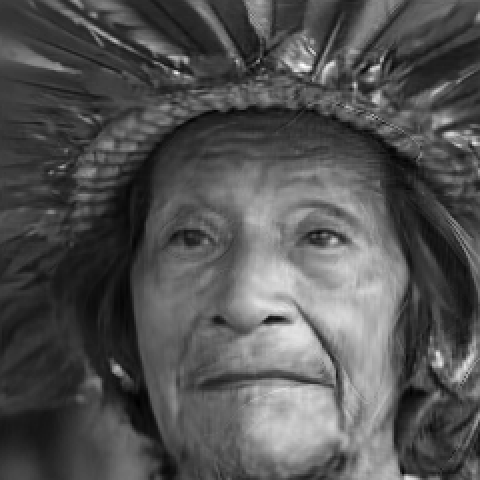} & 
    \includegraphics[width=\sza\columnwidth/2]{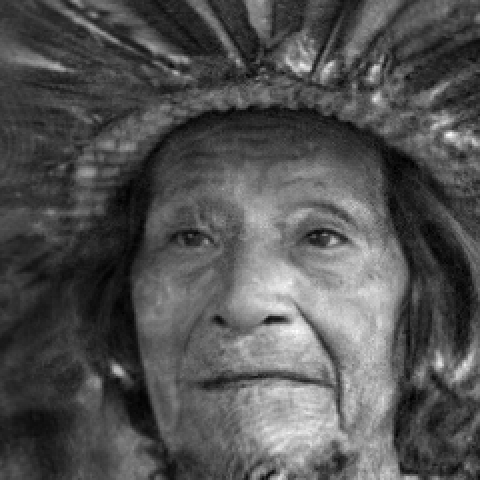} &
    \includegraphics[width=\sza\columnwidth/2]{figs/gt5.png} &
    \hspace{2pt}
    \includegraphics[width=\sza\columnwidth/2]{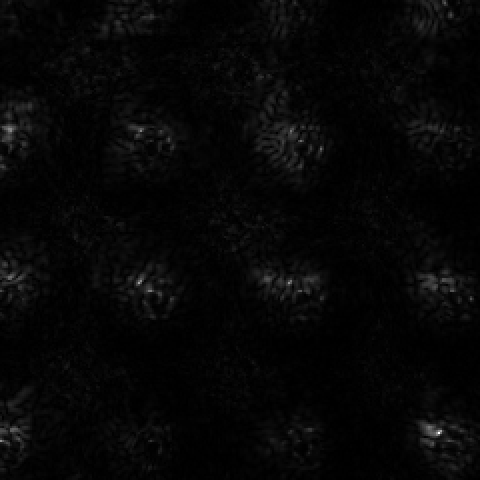} & 
    \includegraphics[width=\sza\columnwidth/2]{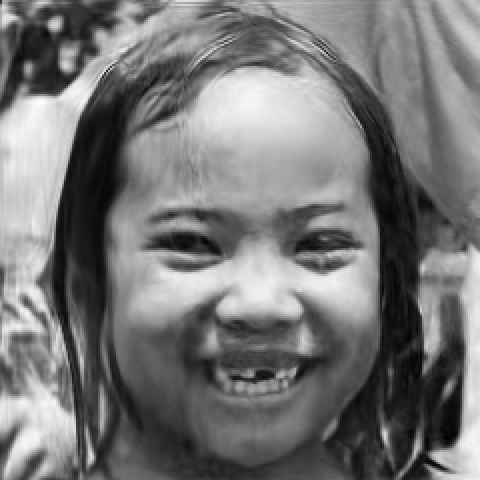} & 
    \includegraphics[width=\sza\columnwidth/2]{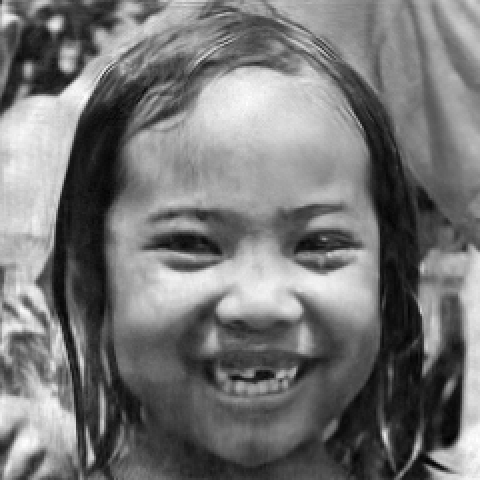} &
    \includegraphics[width=\sza\columnwidth/2]{figs/gt6.png}
    \\
    \rotatebox{90}{\hspace{14pt}\emph{$-50\%$}}
    \hspace{1pt} &
    \includegraphics[width=\sza\columnwidth/2]{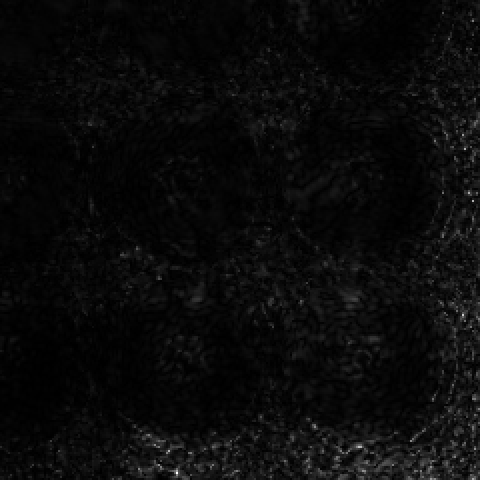} &
    \includegraphics[width=\sza\columnwidth/2]{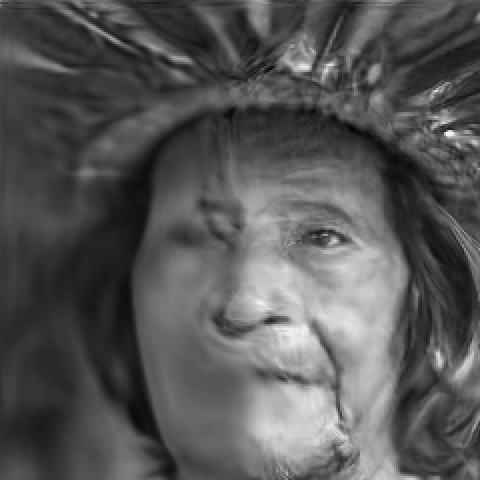} & 
    \includegraphics[width=\sza\columnwidth/2]{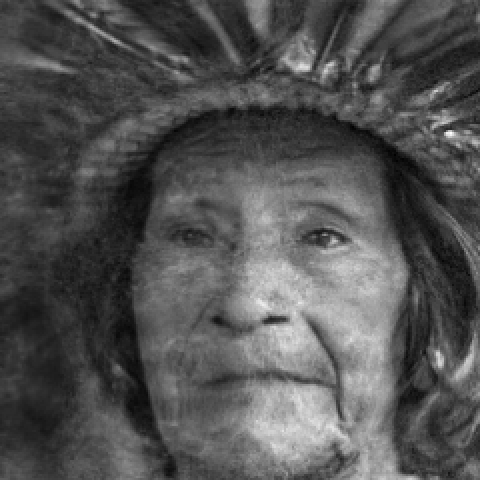} &
    \includegraphics[width=\sza\columnwidth/2]{figs/gt5.png} &
    \hspace{2pt}
    \includegraphics[width=\sza\columnwidth/2]{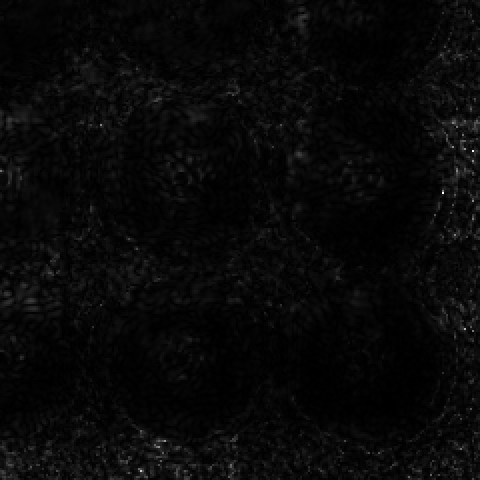} & 
    \includegraphics[width=\sza\columnwidth/2]{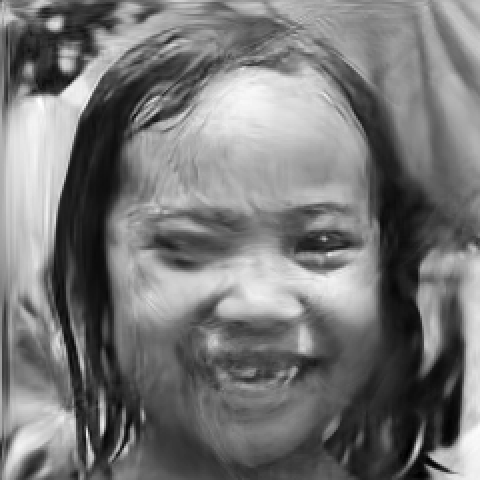} & 
    \includegraphics[width=\sza\columnwidth/2]{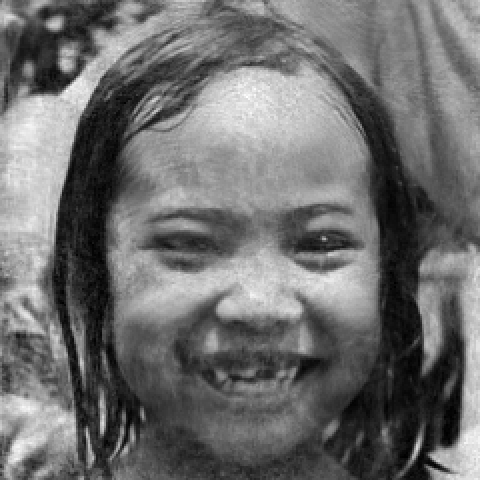} &
    \includegraphics[width=\sza\columnwidth/2]{figs/gt6.png}

\end{tabular}
\caption{Comparison of the reconstructions from the proposed method against ePIE, PtychoGAN, and ground truth for noise-free cases for different overlap percentages.}
 \label{fig:compare_nf}
\end{figure*}

We also compared the performance of the proposed approach under noisy conditions. The approach has two different characteristics in noisy conditions. It has the advantage of having PtychoGAN reconstructions as its initial reconstruction, which performs well under noisy scenarios due to its total variation and deep generative priors. However, it also has the disadvantage of performing ePIE steps as the final part of the algorithm. ePIE does not have any measures to tackle the noise problem in the reconstructions, and if the diffraction data is noisy, the reconstructions tend to fit the noise. 

When we look at Fig.~\ref{fig:compare_n}, we can see the differences in the algorithms. As expected, ePIE cannot perform well under low overlap conditions. Although PtychoGAN handles noise quite successfully, the distortions are worsened with the additive effect of noise in some cases. This is apparent in the low overlap percentage reconstructions. On the other hand, the proposed approach can improve the reconstructions. For the last row for image 1 in Fig.~\ref{fig:compare_n}, the reconstruction shows great improvement in the structure of the face. In addition, the reconstructions show significant improvement in quality outside of the faces, as we have seen in the noise-free cases.

Table~\ref{tab:compare_n_tb} shows the MSE values for the reconstructions in Fig.~\ref{fig:compare_n}. We can see that the performance increase with the proposed approach varies for each case. However, MSE values do not always show  the full significance of the improvements. The case where the proposed approach fixed the structure of the reconstruction for image 1 only decreased the MSE by $1.14^{-3}$ while the visual improvement is more significant than this value. 

\setlength\tabcolsep{3.2pt}
\begin{table}[!b]
\footnotesize
\caption{MSE values for reconstructions in Fig~\ref{fig:compare_n} using different methods with varying overlap ratios.}
\centering
\begin{tabular}{@{}c|x{2cm}|x{2cm}|x{2cm}||x{2cm}|x{2cm}|x{2cm}|@{}}
\multicolumn{1}{|c|}{\textbf{MSE ($x10^{-3}$)}}&
\multicolumn{3}{c||}{\textbf{Image 1}} & \multicolumn{3}{c|}{\textbf{Image 2}} \\
% \cmidrule(l){2-9} 
\midrule
\multicolumn{1}{|c|}{Method} & {ePIE} & {PtychoGAN} & {Proposed}  & {ePIE} & {PtychoGAN} & {Proposed} \\ \midrule
\multicolumn{1}{|c|}{Overlap = $25\%$} & 6.39 & 7.66 & \textbf{4.38} & 61.90 & 2.45 & \textbf{1.35} \\
\multicolumn{1}{|c|}{Overlap = $0\%$} & 68.32 & 1.66 & \textbf{1.43} & 178.79 & 1.81 & \textbf{1.72} \\
\multicolumn{1}{|c|}{Overlap = $-25\%$} & 103.18 & 3.76 & \textbf{2.92} & 233.59 & 7.31 & \textbf{3.78} \\
\multicolumn{1}{|c|}{Overlap = $-50\%$} & 109.95 & 5.71 & \textbf{4.57} & 244.33 & 8.02 & \textbf{4.53} \\
\midrule
\end{tabular}
\label{tab:compare_n_tb}
\end{table}

% \setlength\tabcolsep{3.2pt}
% \begin{table}[!b]
% \footnotesize
% \caption{Caption}
% \centering
% \begin{tabular}{@{}c|x{2cm}|x{2cm}|x{2cm}||x{2cm}|x{2cm}|x{2cm}|@{}}
% \multicolumn{1}{|c|}{\textbf{SSIM}} &
% \multicolumn{3}{c||}{\textbf{Image 1}} & \multicolumn{3}{c|}{\textbf{Image 2}} \\
% % \cmidrule(l){2-9} 
% \midrule
% \multicolumn{1}{|c|}{Method} & {ePIE} & {PtychoGAN} & {Proposed}  & {ePIE} & {PtychoGAN} & {Proposed} \\ \midrule
% \multicolumn{1}{|c|}{Overlap = $25\%$} & 0.726 & 0.860 & \textbf{0.871} & 0.592 & 0.916 & \textbf{0.} \\
% \multicolumn{1}{|c|}{Overlap = $0\%$} & 0.248 & 0.902 & \textbf{0.850} & 0.174 & 0.925 & \textbf{0.} \\
% \multicolumn{1}{|c|}{Overlap = $-25\%$} & 0.112 & 0.843 & \textbf{0.784} & 0.073 & 0.839 & \textbf{0.} \\
% \multicolumn{1}{|c|}{Overlap = $-50\%$} & 0.073 & 0.786 & \textbf{0.711} & 0.055 & 0.795 & \textbf{0.} \\
% \hline
% \end{tabular}
% \label{tab:compare_nf_tb}
% \end{table}

\begin{figure*}[!t]
    \centering
    \setlength{\tabcolsep}{\szb}

    \begin{tabular}{ccccccccc}
    & {ePIE} & {PtychoGAN} & {Proposed} & {GT} 
    & {ePIE} & {PtychoGAN} & {Proposed} & {GT} 
    \vspace{2pt}
    \\
    \rotatebox{90}{\hspace{20pt}\emph{$25\%$}}
    \hspace{1pt} &
    \includegraphics[width=\sza\columnwidth/2]{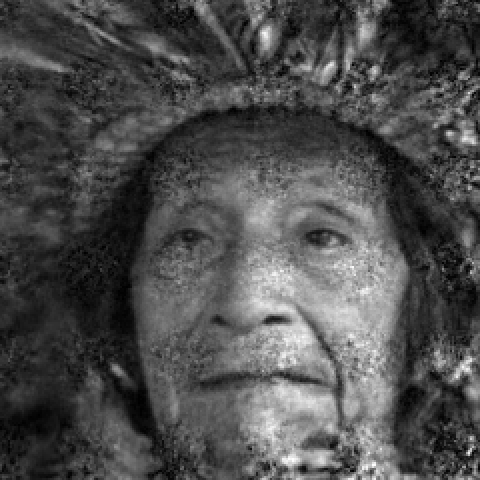} & 
    \includegraphics[width=\sza\columnwidth/2]{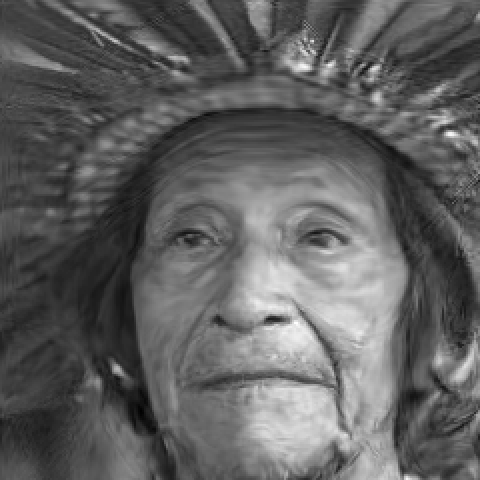} & 
    \includegraphics[width=\sza\columnwidth/2]{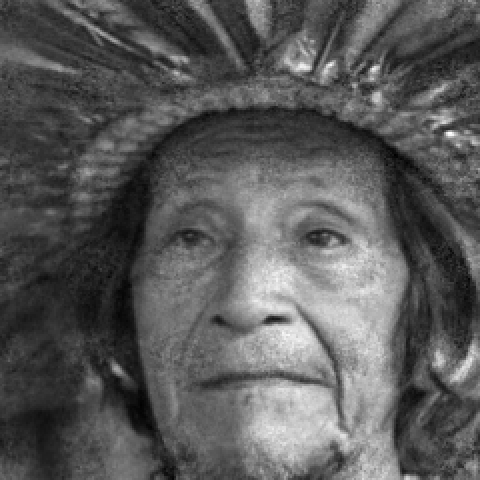} &
    \includegraphics[width=\sza\columnwidth/2]{figs/gt5.png} &
    \hspace{2pt}
    \includegraphics[width=\sza\columnwidth/2]{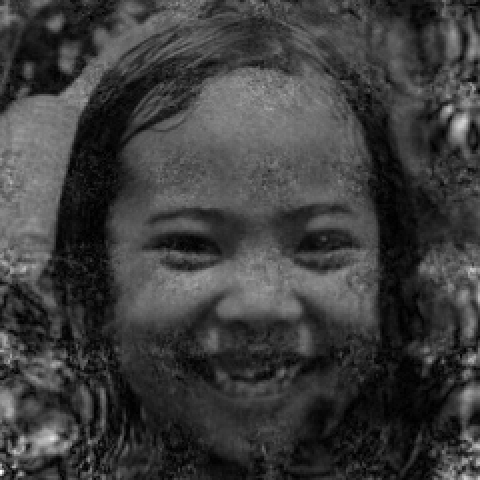} & 
    \includegraphics[width=\sza\columnwidth/2]{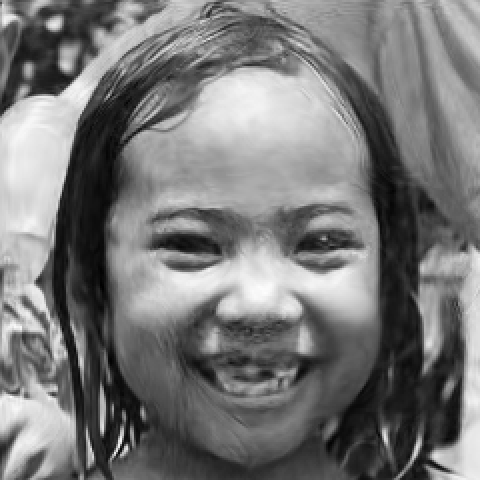} & 
    \includegraphics[width=\sza\columnwidth/2]{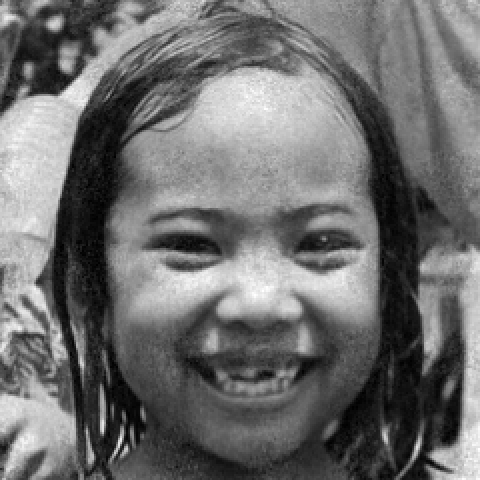} &
    \includegraphics[width=\sza\columnwidth/2]{figs/gt6.png}
    \\
    \rotatebox{90}{\hspace{24pt}\emph{$0\%$}}
    \hspace{1pt} &
    \includegraphics[width=\sza\columnwidth/2]{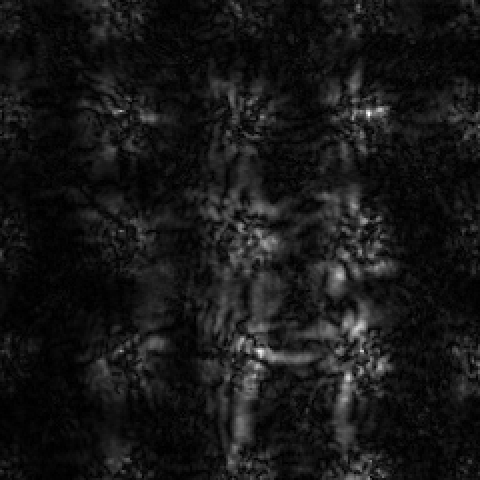} & 
    \includegraphics[width=\sza\columnwidth/2]{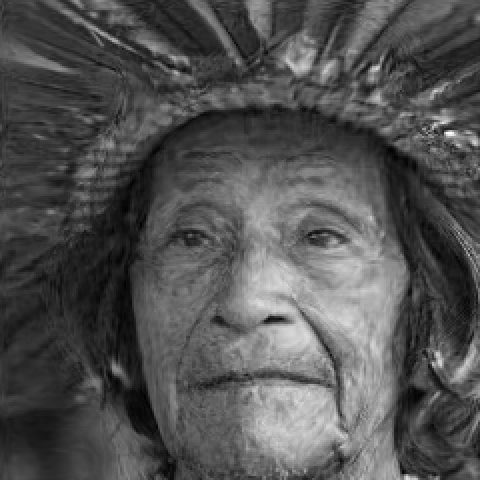} & 
    \includegraphics[width=\sza\columnwidth/2]{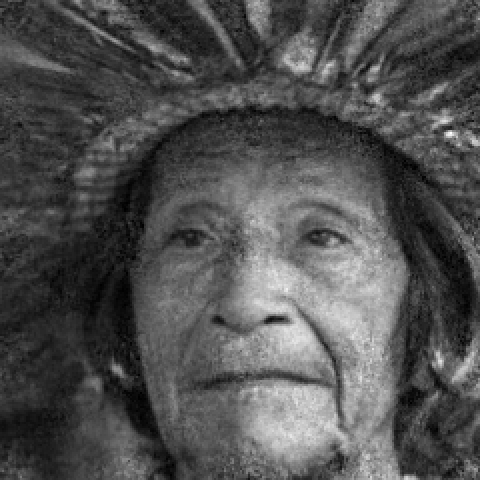} &
    \includegraphics[width=\sza\columnwidth/2]{figs/gt5.png} &
    \hspace{2pt}
    \includegraphics[width=\sza\columnwidth/2]{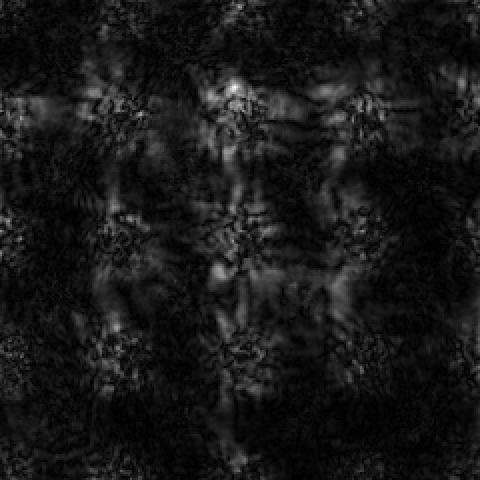} & 
    \includegraphics[width=\sza\columnwidth/2]{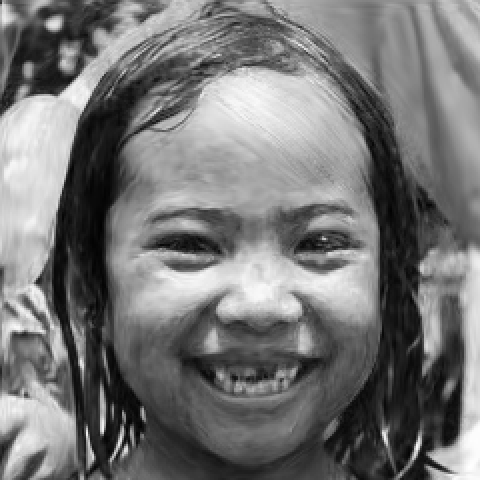} & 
    \includegraphics[width=\sza\columnwidth/2]{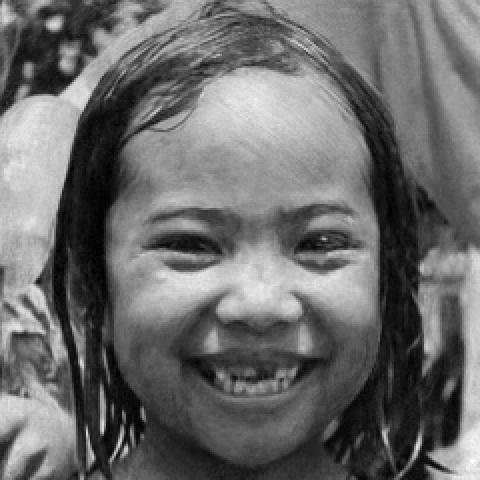} &
    \includegraphics[width=\sza\columnwidth/2]{figs/gt6.png}
    \\
    \rotatebox{90}{\hspace{14pt}\emph{$-25\%$}}
    \hspace{1pt} &
    \includegraphics[width=\sza\columnwidth/2]{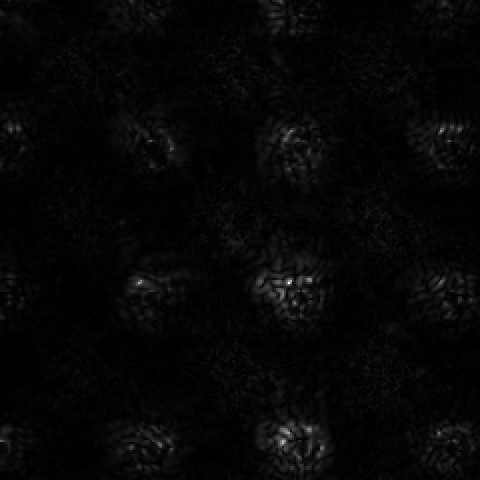} & 
    \includegraphics[width=\sza\columnwidth/2]{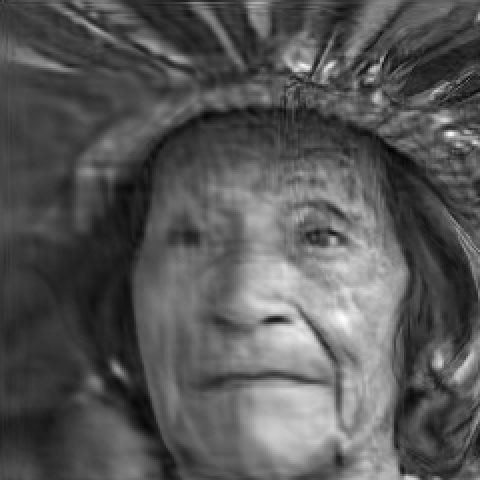} & 
    \includegraphics[width=\sza\columnwidth/2]{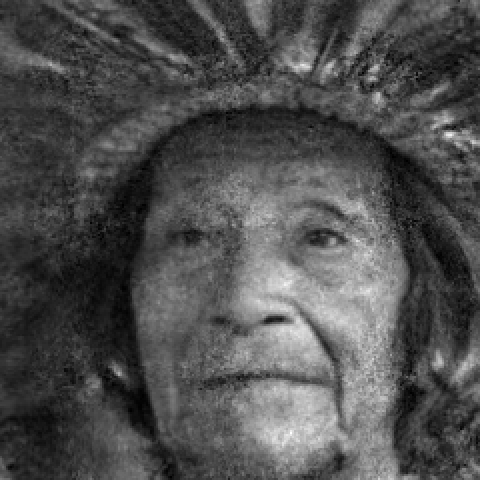} &
    \includegraphics[width=\sza\columnwidth/2]{figs/gt5.png} &
    \hspace{2pt}
    \includegraphics[width=\sza\columnwidth/2]{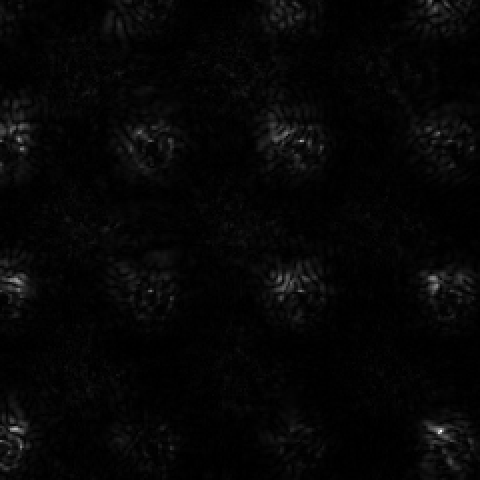} & 
    \includegraphics[width=\sza\columnwidth/2]{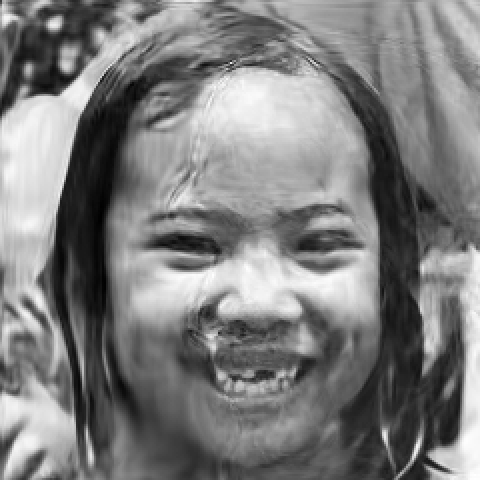} & 
    \includegraphics[width=\sza\columnwidth/2]{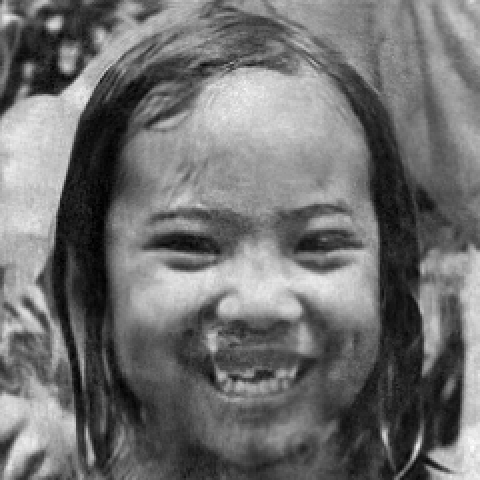} &
    \includegraphics[width=\sza\columnwidth/2]{figs/gt6.png}
    \\
    \rotatebox{90}{\hspace{14pt}\emph{$-50\%$}}
    \hspace{1pt} &
    \includegraphics[width=\sza\columnwidth/2]{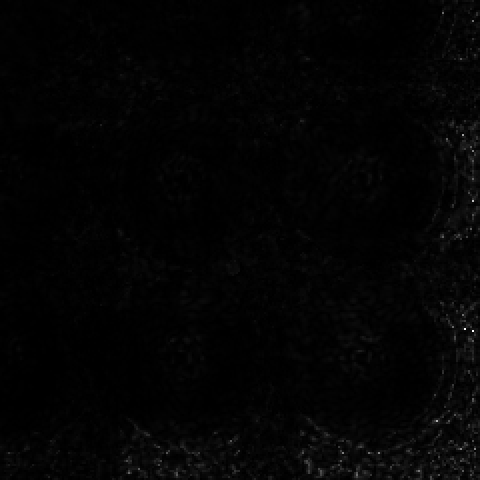} &
    \includegraphics[width=\sza\columnwidth/2]{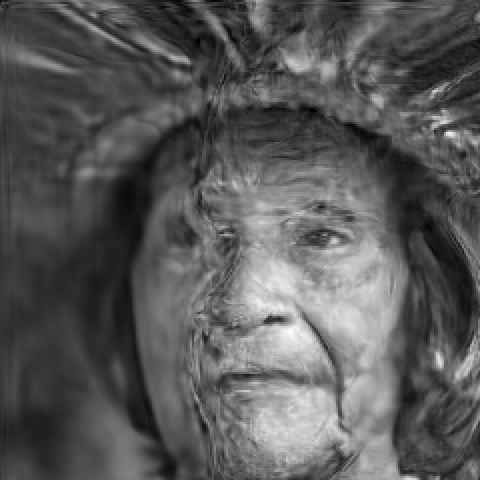} & 
    \includegraphics[width=\sza\columnwidth/2]{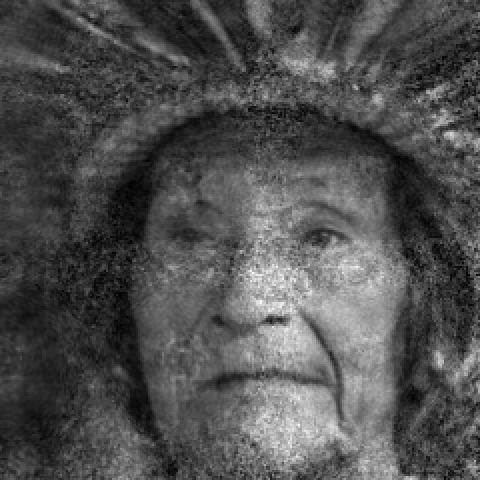} &
    \includegraphics[width=\sza\columnwidth/2]{figs/gt5.png} &
    \hspace{2pt}
    \includegraphics[width=\sza\columnwidth/2]{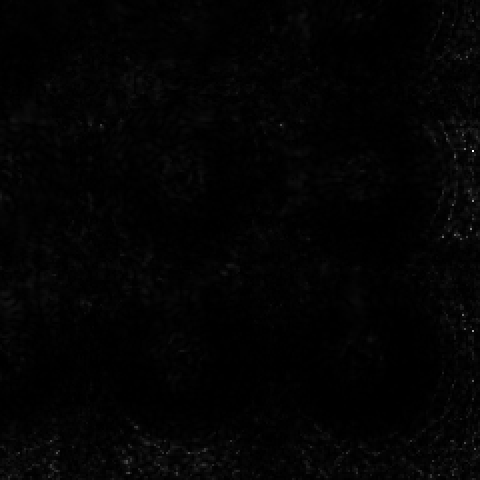} & 
    \includegraphics[width=\sza\columnwidth/2]{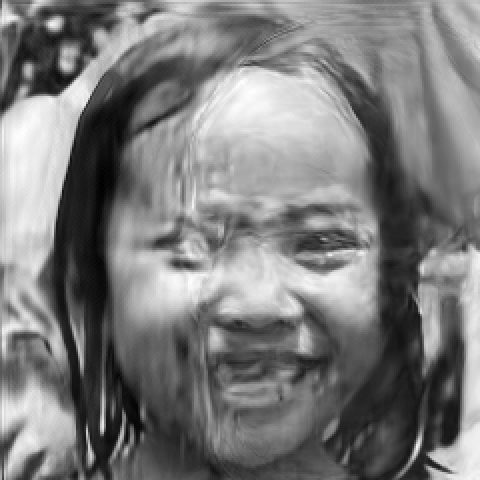} & 
    \includegraphics[width=\sza\columnwidth/2]{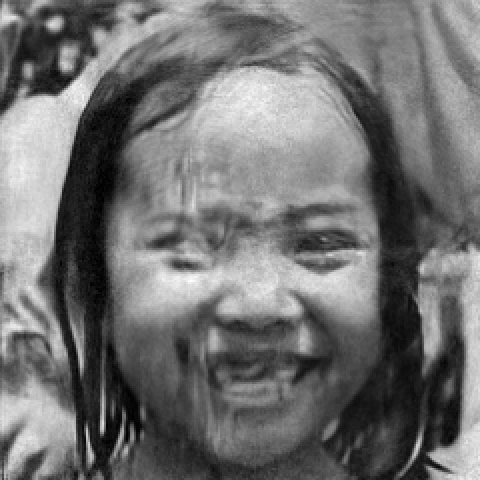} &
    \includegraphics[width=\sza\columnwidth/2]{figs/gt6.png}

\end{tabular}
\caption{Comparison of the reconstructions from the proposed method against ePIE, PtychoGAN, and ground truth for noisy cases for different overlap percentages.}
 \label{fig:compare_n}
\end{figure*}

Finally, we evaluate the performance of the algorithm for an increased number of iterations of the method applied to each method. In Fig.~\ref{fig:compare_nf_multiple}, the reconstruction qualities can be compared qualitatively between single and multiple iterations. Increasing the iteration number increases the fidelity to the data, leading to an improved reconstruction quality. However, increased iterations also amplify the artifacts and noise. This results in a decreased reconstruction quality after some number of iterations, and this decrease happens sooner if there is less overlap in the first place. Thus, in Fig.~\ref{fig:compare_nf_multiple}, results shown for multiple iterations have iteration numbers of 6, 11, 13, and 13, corresponding to $-50\%$, $-25\%$, $0\%$, and $-25\%$ overlap cases, respectively.

Fig.\ref{fig:compare_nf_multiple} and Table~\ref{tab:compare_nf_multiple_tb} show the improvements achieved by the repetition of the algorithm. The reconstruction quality outside the face areas, which are outside of the representation capability of the network, has higher amounts of details with less blur. Similarly, artifacts are reduced significantly. However, in the $-50\%$ overlap case, the noise is also amplified. While quantitative improvements are higher in higher overlap percentages, qualitative improvements in lower overlaps are crucial since visible artifacts are removed. In the case of some overlap, i.e., $25\%$ overlap, the reconstructions become perfect in the absence of noise with an increased iteration number. In addition, the convergence shows improved behavior if $w$ is lowered at each iteration. 

\setlength\tabcolsep{3.2pt}
\begin{table}[!b]
\footnotesize
\caption{MSE values for reconstructions in Fig~\ref{fig:compare_nf_multiple}}
\centering
\begin{tabular}{@{}c|x{2cm}|x{2cm}||x{2cm}|x{2cm}|@{}}
\multicolumn{1}{|c|}{\textbf{MSE ($x10^{-3}$)}}&
\multicolumn{2}{c||}{\textbf{Image 1}} & \multicolumn{2}{c|}{\textbf{Image 2}} \\
% \cmidrule(l){2-9} 
\midrule
\multicolumn{1}{|c|}{\multirow{2}{*}{Method}} & {Single} & {Multiple} & {Single} & {Multiple} 
\\
\multicolumn{1}{|c|}{}& {Iteration} & {Iterations} & {Iteration} & {Iterations}
\\ \midrule
\multicolumn{1}{|c|}{Overlap = $25\%$} & 2.18 & \textbf{0.00} & 0.78 & \textbf{0.00} \\
\multicolumn{1}{|c|}{Overlap = $0\%$} & 3.04 & \textbf{0.03} & 1.04 & \textbf{0.64} \\
\multicolumn{1}{|c|}{Overlap = $-25\%$} & 2.70 & \textbf{0.85} & 2.77 & \textbf{1.71} \\
\multicolumn{1}{|c|}{Overlap = $-50\%$} & 2.89 & \textbf{1.41} & 3.86 & \textbf{3.12} \\
\midrule
\end{tabular}
\label{tab:compare_nf_multiple_tb}
\end{table}

% \setlength\tabcolsep{3.2pt}
% \begin{table}[!b]
% \footnotesize
% \caption{Caption}
% \centering
% \begin{tabular}{@{}c|x{2cm}|x{2cm}|x{2cm}||x{2cm}|x{2cm}|x{2cm}|@{}}
% \multicolumn{1}{|c|}{\textbf{SSIM}} &
% \multicolumn{3}{c||}{\textbf{Image 1}} & \multicolumn{3}{c|}{\textbf{Image 2}} \\
% % \cmidrule(l){2-9} 
% \midrule
% \multicolumn{1}{|c|}{Method} & {ePIE} & {PtychoGAN} & {Proposed}  & {ePIE} & {PtychoGAN} & {Proposed} \\ \midrule
% \multicolumn{1}{|c|}{Overlap = $25\%$} & 0.861 & 0.878 & \textbf{0.961} & 0.668 & 0.898 & \textbf{0.} \\
% \multicolumn{1}{|c|}{Overlap = $0\%$} & 0.256 & 0.863 & \textbf{0.913} & 0.173 & 0.919 & \textbf{0.} \\
% \multicolumn{1}{|c|}{Overlap = $-25\%$} & 0.144 & 0.753 & \textbf{0.876} & 0.080 & 0.885 & \textbf{0.} \\
% \multicolumn{1}{|c|}{Overlap = $-50\%$} & 0.154 & 0.831 & \textbf{0.846} & 0.101 & 0.827 & \textbf{0.} \\
% \hline
% \end{tabular}
% \label{tab:compare_nf_tb}
% \end{table}

\begin{figure*}[!t]
    \centering
    \setlength{\tabcolsep}{\szb}

    \begin{tabular}{ccccccc}
    & {Single} & {Multiple} & \multirow{2}{*}{GT}
    & {Single} & {Multiple} & \multirow{2}{*}{GT}
    \\
    & {Iteration} & {Iterations} & 
    & {Iteration} & {Iterations} & 
    \vspace{2pt}
    \\
    \rotatebox{90}{\hspace{20pt}\emph{$25\%$}}
    \hspace{1pt} &
    \includegraphics[width=\sza\columnwidth/2]{figs/comb/im5/initial45.0to15_noRandShift_l1Loss_weight15_combinedmultiple3.0_discLoss0_realProbe_noisePOISSON0_tvnewmult0_probe120_step15.png} &
    \includegraphics[width=\sza\columnwidth/2]{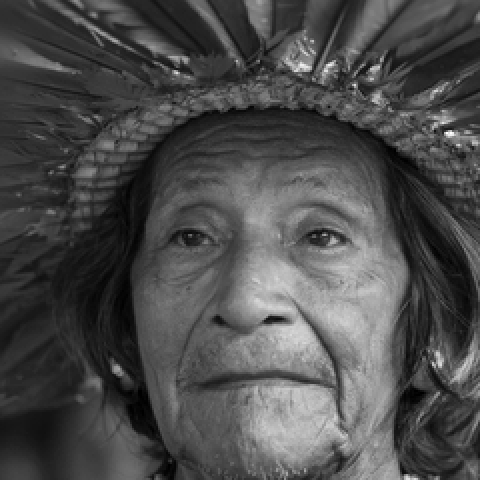} &
    \includegraphics[width=\sza\columnwidth/2]{figs/gt5.png} &
    \hspace{2pt}
    \includegraphics[width=\sza\columnwidth/2]{figs/comb/im6/initial45.0to15_noRandShift_l1Loss_weight15_combinedmultiple3.0_discLoss0_realProbe_noisePOISSON0_tvnewmult0_probe120_step15.png} &
    \includegraphics[width=\sza\columnwidth/2]{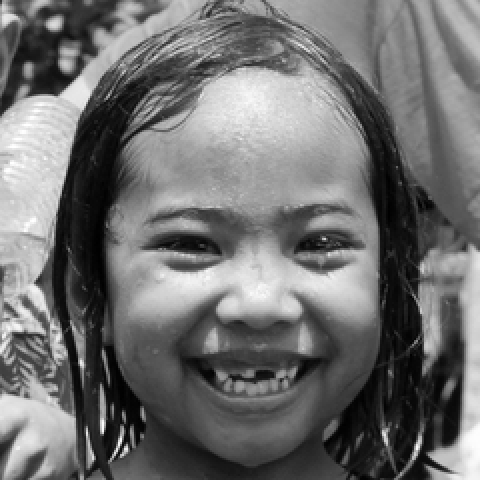} &
    \includegraphics[width=\sza\columnwidth/2]{figs/gt6.png}
    \\
    \rotatebox{90}{\hspace{24pt}\emph{$0\%$}}
    \hspace{1pt} &
    \includegraphics[width=\sza\columnwidth/2]{figs/ratios/im5/initial60.0to20_noRandShift_l1Loss_weight15_combinedmultiple3.0_discLoss0_realProbe_noisePOISSON0_tvnewmult0_probe120_step20.png} &
    \includegraphics[width=\sza\columnwidth/2]{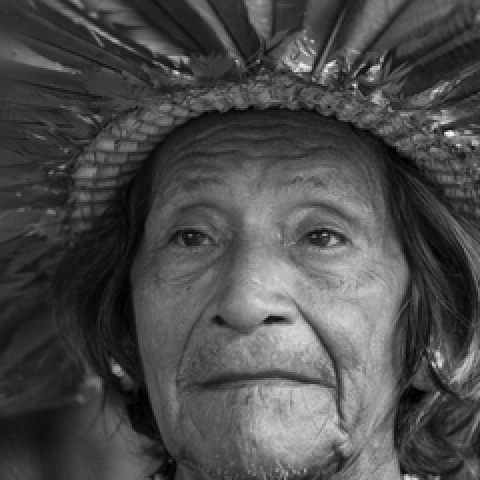} &
    \includegraphics[width=\sza\columnwidth/2]{figs/gt5.png} &
    \hspace{2pt}
    \includegraphics[width=\sza\columnwidth/2]{figs/comb/im6/initial60.0to20_noRandShift_l1Loss_weight15_combinedmultiple3.0_discLoss0_realProbe_noisePOISSON0_tvnewmult0_probe120_step20.png} &
    \includegraphics[width=\sza\columnwidth/2]{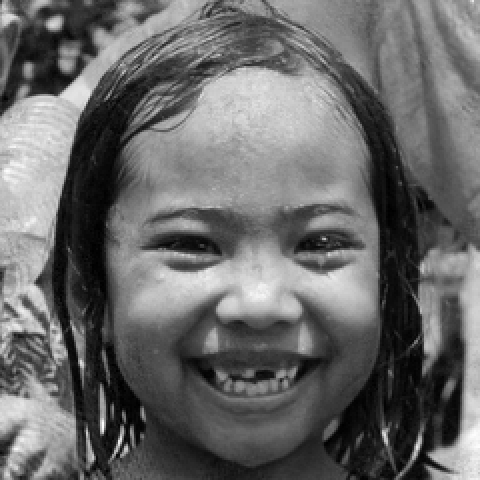} &
    \includegraphics[width=\sza\columnwidth/2]{figs/gt6.png}
    \\
    \rotatebox{90}{\hspace{14pt}\emph{$-25\%$}}
    \hspace{1pt} &
    \includegraphics[width=\sza\columnwidth/2]{figs/comb/im5/initial75.0to25_noRandShift_l1Loss_weight15_combinedmultiple3.0_discLoss0_realProbe_noisePOISSON0_tvnewmult0_probe120_step25.png} &
    \includegraphics[width=\sza\columnwidth/2]{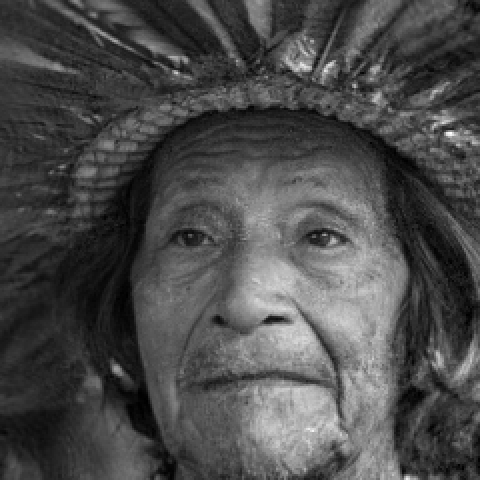} &
    \includegraphics[width=\sza\columnwidth/2]{figs/gt5.png} &
    \hspace{2pt}
    \includegraphics[width=\sza\columnwidth/2]{figs/fixes/initial75.0to15_noRandShift_l1Loss_weight10_combinedmultiple5.0_discLoss0_realProbe_noisePOISSON0_tvnewmult0_probe120_step15.png} &
    \includegraphics[width=\sza\columnwidth/2]{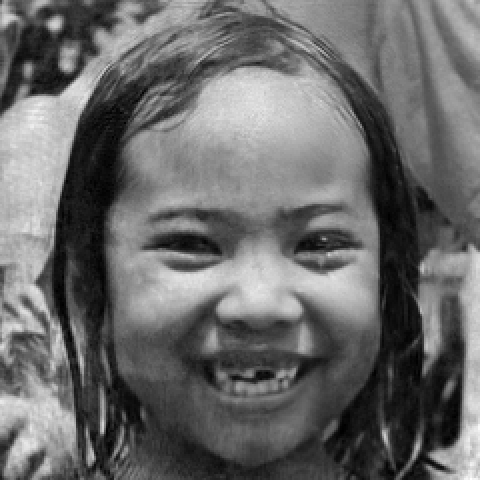} &
    \includegraphics[width=\sza\columnwidth/2]{figs/gt6.png}
    \\
    \rotatebox{90}{\hspace{14pt}\emph{$-50\%$}}
    \hspace{1pt} &
    \includegraphics[width=\sza\columnwidth/2]{figs/comb/im5/initial90.0to30_noRandShift_l1Loss_weight15_combinedmultiple3.0_discLoss0_realProbe_noisePOISSON0_tvnewmult0_probe120_step30.png} &
    \includegraphics[width=\sza\columnwidth/2]{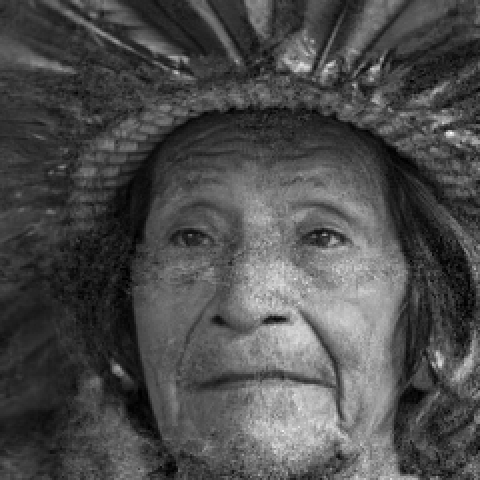} &
    \includegraphics[width=\sza\columnwidth/2]{figs/gt5.png} &
    \hspace{2pt}
    \includegraphics[width=\sza\columnwidth/2]{figs/comb/im6/initial90.0to30_noRandShift_l1Loss_weight15_combinedmultiple3.0_discLoss0_realProbe_noisePOISSON0_tvnewmult0_probe120_step30.png} &
    \includegraphics[width=\sza\columnwidth/2]{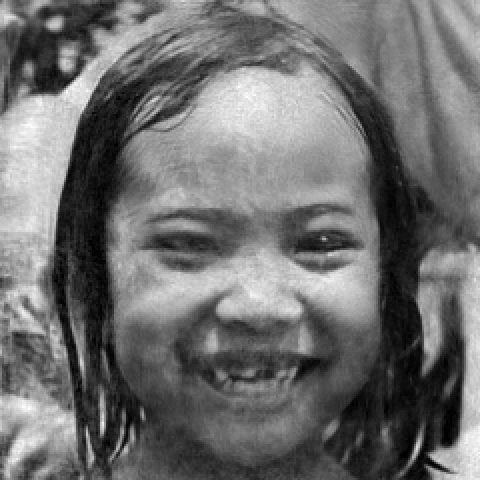} &
    \includegraphics[width=\sza\columnwidth/2]{figs/gt6.png}

\end{tabular}
\caption{Comparison of the reconstructions from the proposed method against the ground truth with single and multiple repetitions.}
 \label{fig:compare_nf_multiple}
\end{figure*}

\section{Discussion and Outlook}

In x-ray ptychography, one of the most challenging problems is the requirement for the high overlap between adjacent scanning positions for high reconstruction quality. The high overlap requirement increases the scanning time significantly, which might not be possible in all experimental setups. Lowering the scanning time can be done by a faster-moving probe, which increases the noise significantly, or by lowering the overlap percentage, which leads to lowered reconstruction quality. In addition, well-known methods in the literature fail to recover the object when there is low or no overlap. Different methods are proposed for tackling this problem; some modify the existing algorithms for increased reconstruction quality, and some apply post-processing to reduce artifacts. However, most of them provide small improvements in the reconstructions. 

With the improvements in the deep learning architectures, improvements have been made in x-ray ptychography. Usage of a combination of DGPs and DIPs \cite{Barutcu22} offered a solution for very sparse scanning ptychography, which was not possible before. The algorithm suggests using a pre-trained network to utilize DGPs, progressively modifying the network weights for DIPs, and using additional priors to lower the artifacts caused by noise to get reconstructions. It is shown to be significantly more successful for low overlap ptychography than well-known methods. 

Although the approach produced significantly improved results, the details of the reconstructions were not perfect due to the generalization effect of DGPs and the highly convex nature of the DIPs network structure. In this paper, we proposed an approach to improve the DGPs - DIPs method by blending the process into an oversampling structure and iterative approach with ePIE steps. This way, we showed that we could increase details in the reconstructions acquired by the previous approach.

The reconstruction quality increase depends on the object and the  diffraction data for the proposed architecture. The reconstruction can show great improvements when the object has areas that the network is not trained on, as shown in Fig.~\ref{fig:compare_nf} and Table~\ref{tab:compare_nf_tb}. Moreover, when the reconstructions show distortions in shape due to the convergence problems of DIPs, the proposed approach can recover the structure by updating the object with ePIE steps which can be seen in Fig.~\ref{fig:compare_n} and Table~\ref{tab:compare_n_tb}. Although the algorithm uses ePIE updates, it does not require a high overlap percentage to converge, thanks to the oversampling process applied. Moreover, the reconstructions are further improved by the application of oversampling multiple times, as shown in Fig.\ref{fig:compare_nf_multiple} and Table~\ref{tab:compare_nf_multiple_tb}, leading to better solutions. 

Despite the improvements in the reconstructions, the approach is computationally expensive since it includes the DGPs - DIPs approach proposed before. The process to optimize the network takes approximately 15 minutes for a $25\%$ overlap case for a single reconstruction. Then the application of the ePIE steps follows with an increased overlap percentage. A similar amount of time is taken for ePIE updates if the oversampling ratio is high. Although this is a significant drawback, the reconstruction time can be lowered significantly by parallel optimization using multiple GPUs. An option to improve the speed significantly is doing low-level operations in different GPUs to overcome the bottleneck \cite{Hidayetolu2019, Majchrowicz2020}. 
\section{Conclusion}

In this paper, we proposed an approach for low or no overlap x-ray ptychography reconstructions. The proposed approach is an improvement on the previously suggested work in \cite{Barutcu22}, where deep image priors and deep generative priors are utilized for acquiring a reconstruction with no overlap data. We demonstrated that by having the initial reconstruction from the previously suggested method and applying oversampling, we could solve the problem using ePIE steps to improve the reconstruction quality. Moreover, we showed that the reconstruction quality could be controlled by changing the oversampling ratio and weighting the object updates based on the origin of the diffraction data used. In addition, we demonstrated that the approach produces crucial differences in noisy cases, which can be helpful for a better reconstruction in some cases. Finally, the reconstructions are shown to be improved significantly by the repetition of the oversampling algorithm with updated data.

% \nocite{*}
\section*{Data and Code Availability}

The data and the code used in this study are available upon request.

\bibliography{7_Bibliography/bibliography.bib}

\section*{Acknowledgements}

This research used resources of the Advanced Photon Source, a U.S. Department of Energy (DOE) Office of Science User Facility at Argonne National Laboratory, and is based on research supported by the U.S. DOE Office of Science-Basic Energy Sciences under contract DE-AC02-06CH11357.

\section*{Author contributions statement}

S.B. implemented the idea and conducted the simulations. S.B., A.K.K., and D.G. conceived the idea, analyzed the results, and wrote the manuscript. All authors reviewed and approved the final version of the manuscript.

\section*{Additional information}

\noindent \textbf{Competing interests:} The authors declare no competing interests.

\end{document}